\documentclass[prl,aps,twocolumn,superscriptaddress]{revtex4}
\usepackage{amsmath,amssymb,graphicx}
\usepackage{pxfonts}
\usepackage{graphicx}
\usepackage{color}
\usepackage{float}
\usepackage{ulem}
\usepackage{siunitx}

\usepackage{hyperref}
\hypersetup{%
	pdftitle={Pattern of indirect excitons in van der Waals heterostructure},%
	pdfauthor={Zhiwen Zhou et al.},%
	pdfpagemode={UseNone},%
	pdfstartview={FitH},%
	breaklinks=true,%
	citecolor=blue,%
	colorlinks=true,%
	linkcolor=blue,%
	urlcolor=blue
}

\begin{document}

\date{\today}

\title{Excitonic pattern formation from a wrinkling instability in a van der Waals heterostructure}

\author{Zhiwen~Zhou}
\affiliation{Department of Physics, University of California San Diego, La Jolla, CA 92093, USA}
\author{L.~H.~Fowler-Gerace}
\affiliation{Department of Physics, University of California San Diego, La Jolla, CA 92093, USA}
\author{W.~J.~Brunner}
\affiliation{Department of Physics, University of California San Diego, La Jolla, CA 92093, USA}
\author{E.~A.~Szwed}
\affiliation{Department of Physics, University of California San Diego, La Jolla, CA 92093, USA}
\author{Michael~M.~Fogler}
\affiliation{Department of Physics, University of California San Diego, La Jolla, CA 92093, USA}
\author{Daniel~E.~Parker}
\affiliation{Department of Physics, University of California San Diego, La Jolla, CA 92093, USA}
\author{L.~V.~Butov} 
\affiliation{Department of Physics, University of California San Diego, La Jolla, CA 92093, USA}

\begin{abstract}
\noindent 
We studied photoluminescence of spatially indirect excitons (IXs) in a MoSe$_2$/WSe$_2$ van der Waals heterostructure. We observed a quasi-periodic triangular pattern of IXs with a characteristic spatial wavelength $\sim{}2.6\,\mu\mathrm{m}$.
Our theoretical analysis using a F\"oppl-von K\'arm\'an theory finds an elastic instability that produces triangular patterns with micrometer-scale wavelengths.
This mechanism is consistent with the wavelength, symmetry, excitation power dependence, and temperature dependence of the observed IX pattern.
\end{abstract}
\maketitle

A spatial modulation of exciton luminescence is a basic phenomenon explored in various excitonic systems. Spatially indirect excitons (IXs), also known as interlayer excitons, are composed of electrons and holes in separated layers~\cite{Lozovik1976}. Due to the layer separation, IX lifetimes are significantly longer than those of regular spatially direct excitons (DXs)~\cite{Zrenner1992}. Their long lifetime allows IXs to cool to low temperatures~\cite{Lozovik1976} and the efficient IX thermalization facilitates a spatial modulation of IX luminescence.

Various mechanisms can drive modulations in excitonic systems. For instance, a periodic modulation of exciton luminescence can be caused by a Turing instability driven by stimulated kinetics of exciton formation due to quantum degeneracy~\cite{Levitov2005, Levitov2005a}. Studies of IXs in GaAs heterostructures led to the observation of the exciton ring modulation~\cite{Butov2002, Butov2004, Yang2015}, which is caused by this instability~\cite{Levitov2005}. The wavelength of this instability is controlled by density~\cite{Levitov2005} and varies in the range $9$--$40\,\mu\mathrm{m}$ in the experiments~\cite{Butov2002, Butov2004, Yang2015}. A triangular pattern of IXs due to this instability was predicted for electron-hole systems~\cite{Levitov2005a}.

A modulation of exciton luminescence can also emerge due to attractive interaction. For instance, a modulation due to attractive interaction was observed in cold atoms~\cite{Strecker2002}. 

Van der Waals heterostructures allow exploring new mechanisms for modulations of exciton luminescence. Twisting between the layers in a van der Waals heterostructure can lead to IX modulation caused by moir{\'e}~\cite{Wu2018, Yu2018, Wu2017, Yu2017, Zhang2017a, Zhang2018, Ciarrocchi2019, Seyler2019, Tran2019, Jin2019, Alexeev2019, Jin2019a, Shimazaki2020, Wilson2021, Gu2022} and super moir{\'e}~\cite{Wang2019, Xia2025, Xie2025} lattices. The modulation period is given by the period of the moir{\'e} superlattice and is typically in the range from few nm to few hundred nm~\cite{Wu2018, Yu2018, Wu2017, Yu2017, Zhang2017a, Zhang2018, Ciarrocchi2019, Seyler2019, Tran2019, Jin2019, Alexeev2019, Jin2019a, Shimazaki2020, Wilson2021, Gu2022, Wang2019, Xia2025, Xie2025}.

In addition to exploring spatially modulated excitonic patterns, it is also essential to explore implications the excitonic patterns may have for exciton transport, collective effects, or excitonic phases. For instance, due to their long lifetimes IXs give an opportunity to realize long-range exciton transport. Consequently, IX transport in van der Waals heterostructures is intensively studied~\cite{Rivera2016, Jauregui2019, Unuchek2019a, Unuchek2019, Liu2019, Choi2020, Huang2020, Yuan2020, Li2021, Wang2021, Shanks2022, Sun2022, Tagarelli2023, Rossi2023, Gao2024, Zhang2024, Wietek2024, Fowler-Gerace2021, Peng2022, Troue2023, Cutshall2025, Fowler-Gerace2024, Zhou2024}. It is essential to verify what implications the excitonic patterns may have for IX transport in van der Waals heterostructures.

In this work, we present the observation of a quasi-periodic triangular pattern of IXs with a characteristic spatial wavelength $\sim 2.6\,\mu\mathrm{m}$ in a MoSe$_2$/WSe$_2$ van der Waals heterostructure. We compare the observed pattern with various modulation mechanisms. We show that elastic instabilities can produce triangular patterns with micrometer-scale  wavelengths in van der Waals heterostructures. This mechanism is consistent with the wavelength, symmetry, excitation power dependence, and temperature dependence of the observed IX pattern.

\begin{figure*}
\begin{center}
\includegraphics[width=14.5cm]{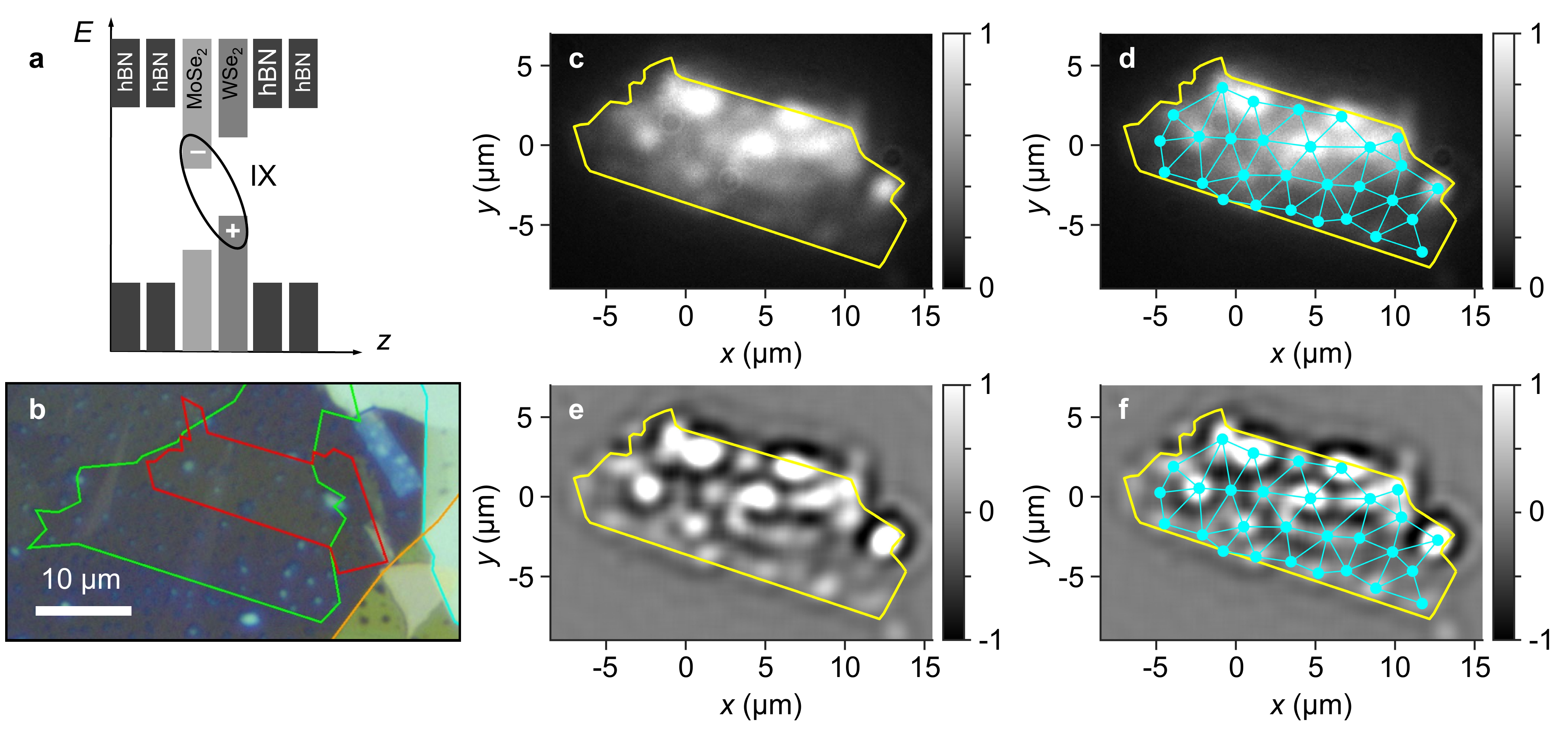}
\caption{IX pattern. 
(a) Schematic energy-band diagram for the heterostructure. The oval indicates an indirect exciton (IX) composed of an electron ($-$) and a hole ($+$).
(b) A microscope image showing the layers of the heterostructure. Scale bar is 10~$\mu$m. The red, green, cyan, and orange lines indicate the boundaries of MoSe$_2$ and WSe$_2$ monolayers and bottom and top hBN layers, respectively.
(c,d) An image of IX PL intensity $I(x,y)$. 
(e,f) The pattern of $- \Delta I(x,y)$ highlighting the spatial modulation of $I(x,y)$. The cyan dots in (d,f) indicate the positions of local maxima in $- \Delta I(x,y)$. These maxima form a quasi-periodic triangular pattern. 
The yellow line in (c-f) shows the boundary of the MoSe$_2$/WSe$_2$ heterostructure.
The laser excitation power $P_\mathrm{ex} = 0.2\,\mathrm{mW}$, temperature $T = 1.7$~K. The $\sim 2\,\mu\mathrm{m}$ laser excitation spot is centered at $x = 1.2\,\mu\mathrm{m}$,~$y = 2.7\,\mu\mathrm{m}$. 
}
\end{center}
\label{fig:spectra}
\end{figure*}

\begin{figure}
\begin{center}
\includegraphics[width=8.5cm]{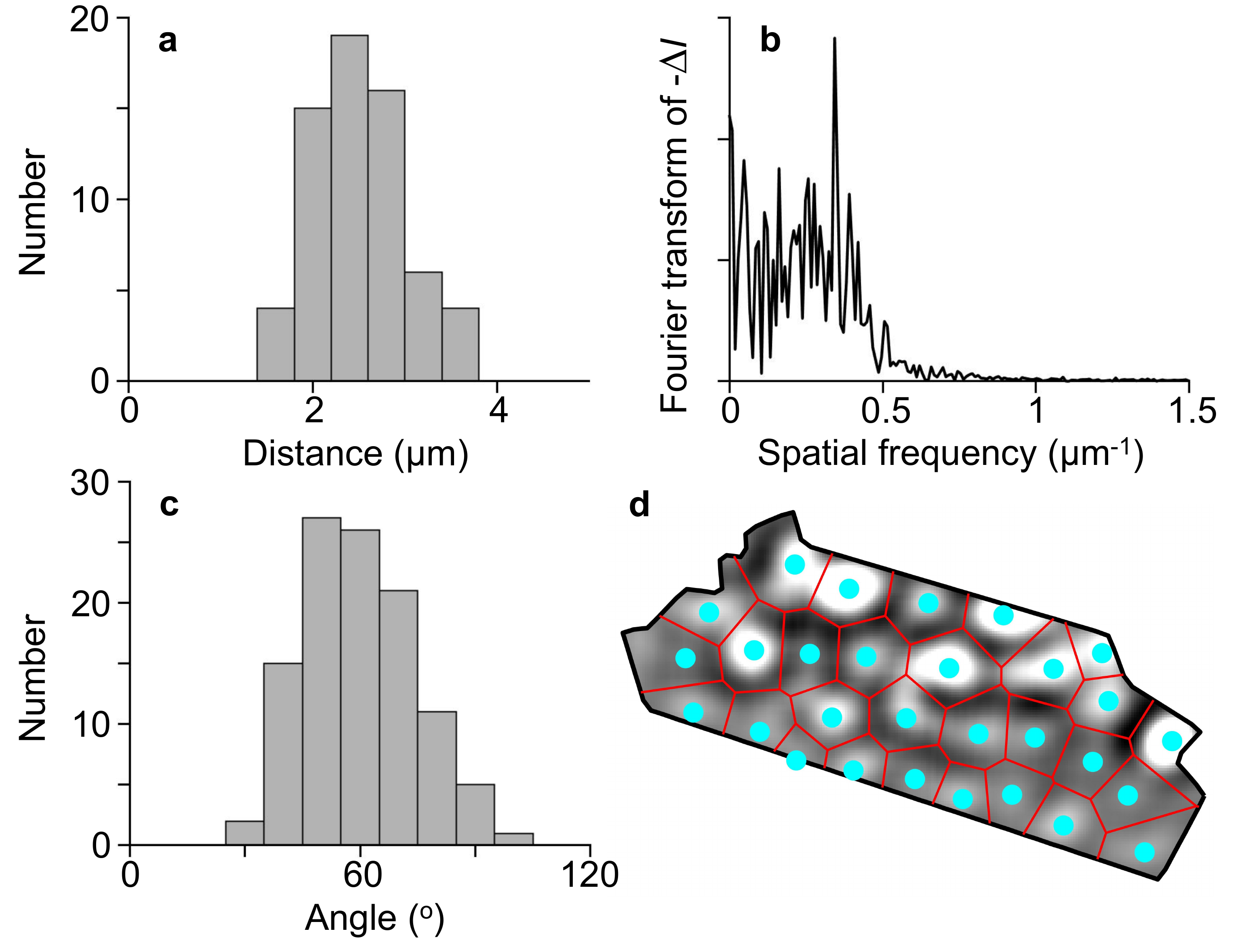}
\caption{Characteristics of IX pattern. 
(a) A histogram of the distances between neighboring local maxima in $-\Delta I(x,y)$. The characteristic wavelength $\lambda \sim 2.6\,\mu\mathrm{m}$. 
(b) The Fourier transform of $-\Delta I(x,y)$ along the line connecting the local maxima. The broad peak on a noise background corresponds to a quasi-periodic modulation with the $\lambda$ lengthscale.
(c) A histogram of the angles between the lines connecting the neighboring local maxima in $-\Delta I(x,y)$. The characteristic angle is $\sim 60^{\circ}$. 
(d) The Voronoi diagram for the local maxima in $-\Delta I(x,y)$ shown by cyan dots. The Voronoi cells are marked by red lines. The average coordination number for the full Voronoi cells, unbroken by the heterostructure edges, is six. 
The data correspond to the IX pattern in Fig.~1.
These data characterize the IX pattern as a distorted triangular pattern. 
}
\end{center}
\label{fig:spectra}
\end{figure}

\begin{figure*}
\begin{center}
\includegraphics[width=14cm]{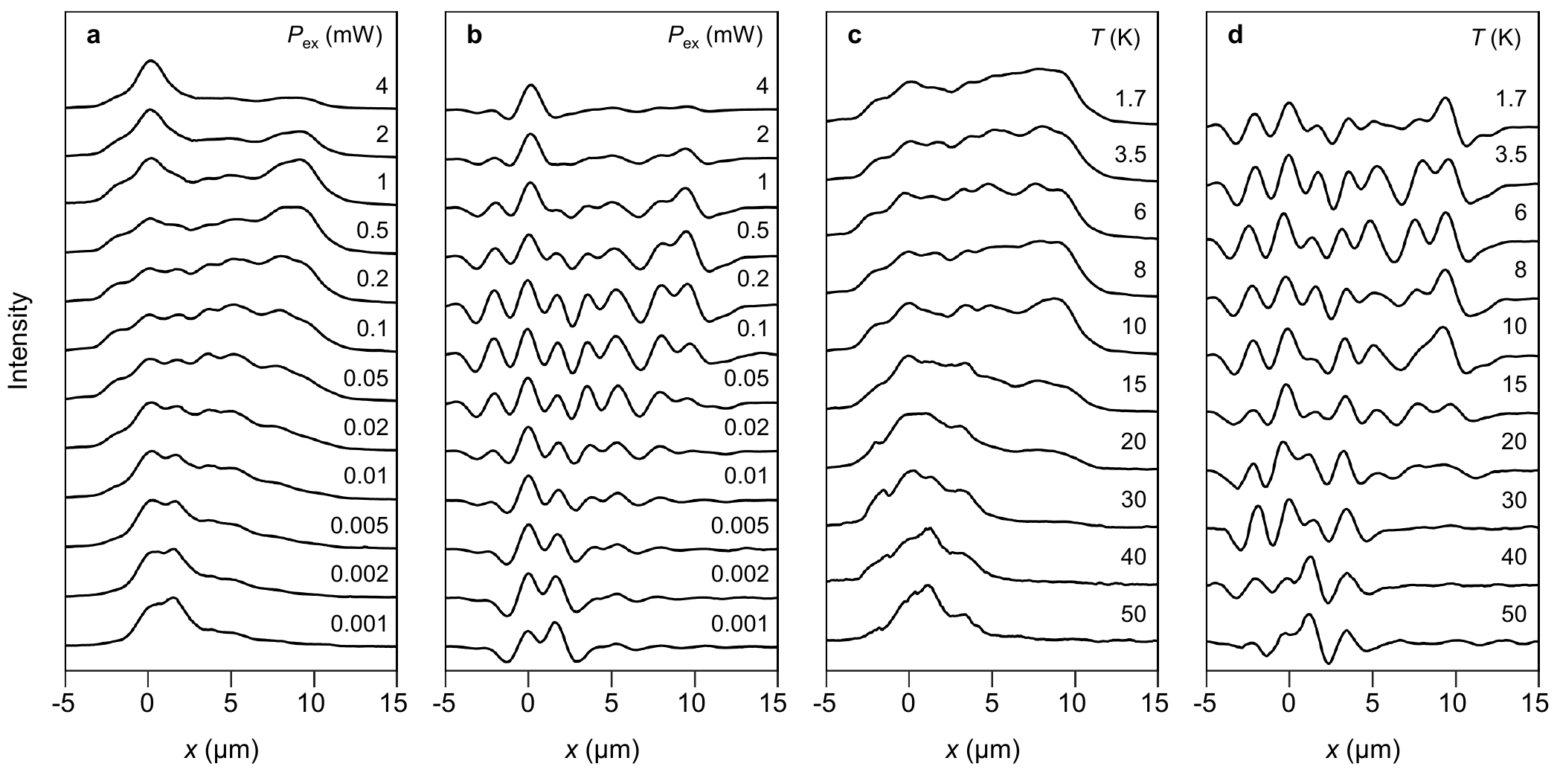}
\caption{Density and temperature dependence. (a-d) Normalized profiles $I(x)$ (a,c) and $- \Delta I(x)$ (b,d) of IX PL showing the quasi-periodic modulation of $I(x)$ (a,c) and $- \Delta I(x)$ (b,d) along $x$ for different excitation densities $P_\mathrm{ex}$ (a,b) and temperatures $T$ (c,d). 
The $\sim 2\,\mu\mathrm{m}$ laser excitation spot is centered at $x = 0$, $y = 0$. $T = 3.5\,\mathrm{K}$ in (a,b), $P_\mathrm{ex} = 0.2\,\mathrm{mW}$ in (c,d).
}
\end{center}
\label{fig:spectra}
\end{figure*}

\vskip5mm \textbf{Experiments.} We study IXs in a MoSe$_2$/WSe$_2$ heterostructure (Fig.~1a) where adjacent MoSe$_2$ monolayer and WSe$_2$ monolayer form the separated electron and hole layers for IXs~\cite{Rivera2015}. The heterostructure was assembled as described in Ref.~\cite{Fowler-Gerace2024} where the same heterostructure was used for IX transport studies. The dry-transfer peel technique~\cite{Withers2015} was used. The MoSe$_2$ and WSe$_2$ monolayers are encapsulated by hBN (Fig.~1a,b) with $\sim 40\,\mathrm{nm}$-thick bottom hBN and $\sim 30\,\mathrm{nm}$-thick top hBN covering the entire MoSe$_2$ and WSe$_2$ layers. The long WSe$_2$ and MoSe$_2$ edges were used for a rotational alignment between the WSe$_2$ and MoSe$_2$ monolayers. The twist angle between these monolayers $\delta \theta = 1.1^\circ$ corresponding to the moir\'e superlattice period $b \sim a / \delta \theta \sim 17\,\mathrm{nm}$ ($a$ is the lattice period) agrees with the angle seen between the long WSe$_2$ and MoSe$_2$ edges in the heterostructure~\cite{Fowler-Gerace2024}. The moir\'e potentials can be affected by atomic reconstruction~\cite{nam2017lattice, Weston2020, Rosenberger2020, Zhao2023} and disorder and may vary over the heterostructure area. The IX $g$-factor $g \approx -16$ in the heterostructure corresponds to the H stacking~\cite{Seyler2019, Wozniak2020}. The mechanically exfoliated layers of the heterostructure were stacked on graphite substrate pressed onto Si/SiO$_2$ substrate.

Excitons were generated by a cw Ti:Sapphire laser with the excitation energy $E_{\rm ex} = 1.689\,\mathrm{eV}$ resonant to DX in WSe$_2$ heterostructure layer. The laser excitation was focused to $\sim 2\,\mu\mathrm{m}$ spot. The IX photoluminescence (PL) was selected by a filter $E \lesssim 1.46\,\mathrm{eV}$ and images of IX PL intensity were measured using a liquid-nitrogen-cooled CCD with a spatial resolution of $0.8\,\mu\mathrm{m}$. The experiments were performed in a variable-temperature $^4$He cryostat. The IX pattern reported in the paper was reproducible after many cycles of cooling down to $\sim 2\,\mathrm{K}$ and warming up to room temperature.

The IX PL images (Fig.~1c) show a spatial modulation of IX PL intensity $I(x,y)$. This modulation is more pronounced in Fig.~1e showing $-\Delta I$ where $\Delta = \partial_x^2 + \partial_y^2$ is the in-plane Laplacian. The cyan dots in Figs.~1d and 1f indicate the positions of local maxima in $-\Delta I(x,y)$. These maxima form a quasi-periodic IX pattern.

The characteristics of this IX pattern are presented in Fig.~2. The average distance between neighboring local maxima in $-\Delta I(x,y)$ is $\sim 2.6\,\mu\mathrm{m}$ (Fig.~2a). This lengthscale corresponds to the broad peak in the Fourier transform of $- \Delta I(x,y)$ along the line connecting the local maxima (Fig.~2b). The characteristic angle between the lines connecting the neighboring local maxima in $-\Delta I(x,y)$ is $\sim 60^{\circ}$ (Fig.~2c). Figure~2d shows the Voronoi tessellation for the local maxima in $-\Delta I(x,y)$. For the Voronoi tessellation, the number of neighbors that share a common edge gives the coordination number. The Voronoi tessellation shows that for nine full Voronoi cells away from the edges, seven have six, one has five, and one more has seven neighbors. The average coordination number is six. These data characterize the IX pattern as a distorted triangular lattice with the period $\lambda \sim 2.6\,\mu\mathrm{m}$.

The power density and temperature dependence of the IX modulation are presented in Fig.~3 showing the $I(x)$ and $-\Delta I(x)$ profiles along $x$. The IX PL signal is integrated within $1\,\mu\mathrm{m}$ in the $y$ direction. The IX transport from the laser excitation spot enhances with increasing laser excitation power $P_\mathrm{ex}$ for $P_\mathrm{ex} \lesssim 0.2\,\mathrm{mW}$ and reduces with increasing $P_\mathrm{ex}$ for higher $P_\mathrm{ex}$, which is seen from the variation of the IX decay distance in Fig.~3a,b. The IX long-range transport vanishes at temperatures above $\sim 10$~K (Fig.~3c,d). These variations of IX transport with the density and temperature were studied in Ref.~\cite{Fowler-Gerace2024}. Figure~3 shows that the quasi-periodic IX modulation is observed along the entire IX signal extension for all the densities and temperatures and the modulation wavelength essentially does not change with density or temperature. 

The efficient IX transport with IXs travelling from the laser excitation spot ($x = 0$) to the heterostructure edge ($x \sim 10\,\mu\mathrm{m}$) with essentially no decay is observed at intermediate densities and low temperatures in Fig.~3. This efficient IX transport, studied in Ref.~\cite{Fowler-Gerace2024}, coexists with the quasi-periodic IX modulation (Fig.~3). 

Figure~3 shows the difference in intensity of the PL maxima. The variation of the intensities of the maxima (Fig.~3), along with the variation of the distances between the maxima (Fig.~2a) and the variation of the angles between the lines connecting the maxima (Fig.~2c), contribute to the deviation of the IX pattern from a perfect triangular pattern. To note the imperfections, we describe the IX pattern as distorted or quasi-periodic triangular pattern.

\vskip5mm \textbf{Structural Mechanism for the Quasi-periodic Pattern.} We now investigate the origin of the quasi-periodic triangular pattern. We first demonstrate that several common mechanisms are disfavored by the data, then show that the quasi-periodic pattern may be produced by buckling of the encapsulating hBN layers residing on the graphite substrate. 

A periodic modulation of exciton PL caused by a Turing instability (modulational instability) driven by kinetics of exciton formation, mentioned in the introduction, vanishes with increasing temperature and the modulation wavelength changes with density~\cite{Levitov2005, Levitov2005a, Butov2002, Butov2004, Yang2015}. However, the IX modulation is observed along the IX signal extension at all temperatures (Fig.~3c,d), and the IX modulation wavelength essentially does not change with the density (Fig.~3a,b). Therefore, the Turing instability mechanism for the observed IX pattern can be ruled out.

IXs are dipoles oriented perpendicular to the heterostructure layers and interaction between them is dominated by the dipolar repulsion as shown by the studies of IX interaction both in GaAs heterostructures~\cite{Yoshioka1990, Butov1994, Zhu1995, Lozovik1997, Palo2002, Ivanov2002, Leon2003, Schindler2008, Laikhtman2009, Maezono2013, Remeika2015} and in van der Waals heterostructures~\cite{Rivera2015, Nagler2017, Kremser2020, Sun2022, Fowler-Gerace2024}. Therefore, the attractive-interaction mechanism of instability~\cite{Strecker2002} is also an unlikely explanation for the observed IX pattern. 

The characteristic period of the pattern $\lambda \sim 2.6\,\mu\mathrm{m}$ is significantly larger than the moir{\'e} period $\sim 17\,\mathrm{nm}$~\cite{Fowler-Gerace2024} due to the lattice mismatch/misalignment of the MoSe$_2$ and WSe$_2$ layers. Therefore, the moir\'e superlattice formation does not explain the observed IX pattern either.

Our sample contains structural imperfections referred to as bubbles. With a few exceptions, the positions of the bubbles do not correlate with the positions of the IX PL maxima (see Supplemental Information, SI, Fig.~S3), also disfavoring them as the origin of the observed pattern. Graphite and hBN are strongly bonded away from the bubbles, evidenced by the reproducibility and lack of degradation in the IX transport after multiple thermal cycles between $T \sim 2\,\mathrm{K}$ and room temperature.

\begin{figure}[h]
\centering
\includegraphics[width=\linewidth]{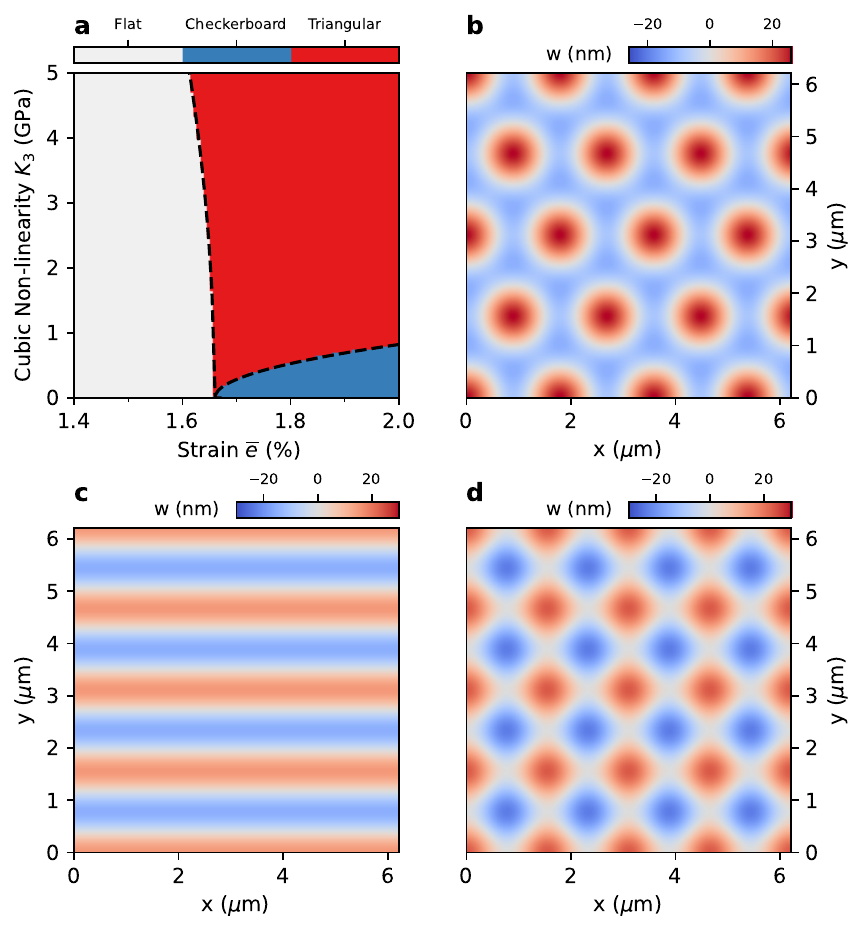}
\caption{
Buckling patterns from the theoretical model.
(a) Phase diagram as a function of the coupling constant $K_3$ and the compressive strain $\bar{e}$ in the pre-buckled state.
The boundaries of the triangular phase (dashed lines) are given by 
$K_{3} = 0.63 E^* ( e_\mathrm{c} - \bar{e})^{1/2} / e_\mathrm{c}$
and
$K_{3} = 0.038 E^* (\bar{e} - e_\mathrm{c})^{1/2} / e_\mathrm{c}$.
(b-d) Optimal buckling profiles for the triangular, stripe, and checkerboard phases at $\bar{e} = 1.67\%$
and $K_3 = 1\,\mathrm{GPa}$. Note that the stripe phase (panel c) is never favored.
Parameters: $h = 70\,\mathrm{nm}$, 
$\overline{E}_\mathrm{f} = 811\,\mathrm{GPa}$, $E^* = 6.14\,\mathrm{GPa}$.
}
\label{fig:FvK_buckling_patterns}
\end{figure}

Finally, as proposed above, the quasi-periodic pattern can emerge due to a structural transition. It is known that when a stiff film bonded to a soft substrate is subjected to a (pre)strain $\bar{e}$ above a critical threshold, it undergoes a buckling transition to a phase with periodic wrinkles~\cite{Allan1969}. Various wrinkle patterns have been studied, including stripe, checkerboard, Miura-ori (or herringbone), and triangular (hexagonal) ones~\cite{Tanaka1987, Bowden1998, Chen2004, Mahadevan2005, Huang2005, Guvendiren2009, Breid2009, Breid2011, Cai2011, Chen2014}.
In these works, isotropic strain was typically imposed by swelling or thermal expansion of the film. Recently, 
wrinkling phenomena have been explored in 2D materials~\cite{Bao2009, Berry2016, Iguiniz2019, Thi2020, Ares2021, Aditya2023}, where the strain often appears during sample fabrication.

The analytical theory of the buckling transition is based on treating the ratio $E^* / \overline{E}_\mathrm{f} \ll 1$ as a small parameter. Here $E^*$ is the effective elastic modulus (also known as the indentation modulus~\cite{Gross2013}) of the substrate.
$\overline{E}_\mathrm{f} = E_\mathrm{f} / (1 - \nu_\mathrm{H,f}^2)$ is the combination the film's Young modulus $E_\mathrm{f}$ and in-plane Poisson ratio $\nu_\mathrm{H,f}$ that determines its bending stiffness $D = \overline{E}_\mathrm{f} h^3 / 12$, with $h$ being the film thickness. 
The pyrolytic graphite in our sample can be treated as a soft substrate for hBN.
Its effective elastic modulus is (see SI) the geometric average of the out-of-plane modulus $C_{33}$ and the shear modulus $C_{44}$:
\begin{equation}
E^* = 2 \sqrt{C_{33} C_{44}}.
\label{eqn:E*}
\end{equation}
For pyrolytic graphite,
the value of $C_{44} \approx \SI{5.0}{GPa}$~\cite{Bosak2007} in single crystals is reduced by one or two orders of magnitude when mobile dislocations are present~\cite{Kelly1981}. With $C_{44} = 0.26\,\mathrm{GPa}$ per~\cite{Blakslee1970}, Eq.~\eqref{eqn:E*} yields $E^* = 6.14\,\mathrm{GPa}$, which we use in our calculations. Independent nanoindentation measurements of pyrolytic graphite reported $E^* = 7.5\text{-}28\,\mathrm{GPa}$, see~\cite{Gross2013} and references therein.
Since the elastic modulus of hBN is $\overline{E}_\mathrm{f} = 811\,\mathrm{GPa}$~\cite{Bosak2006}, our system is indeed in the regime $E^* / \overline{E}_\mathrm{f} \ll 1$.

The theory of the buckling instability predicts that once the strain exceeds a critical threshold, the sample will wrinkle with a characteristic wavelength~\cite{Allan1969}
\begin{equation}
\lambda_\mathrm{w} = 2\pi h \left( \frac{\overline{E}_\mathrm{f}}{3 E^*} \right)^{1/3} . 
\label{eqn:lambda_w}
\end{equation}
Additional strain beyond criticality changes $\lambda_{\mathrm{w}}$ only marginally, see~\cite{Bowden1998, Mahadevan2005, Chen2014} and SI.
For $h = 70\,\mathrm{nm}$, Eq.~\eqref{eqn:lambda_w} predicts that the interpeak distance for the triangular wrinkle pattern
is $({2} /\!{\sqrt{3}}\,) \lambda_{\mathrm{w}} \sim \SI{1.9}{\micro\meter}$,
which is comparable to the observed interpeak distance $\lambda \sim 2.6\,\mu\mathrm{m}$ of the IX PL.

To understand the wrinkle pattern quantitatively, we employed the F\"oppl-von K\'arm\'an (FvK) plate theory,
which is commonly used to study buckling of thin films on compliant substrates~\cite{Chen2004, Huang2005, audolyboudaoud2008part1, Cai2011}.
The FvK energy density functional~\cite{landau2012theory} describes bending and stretching of the film:
\begin{equation}
\begin{aligned}
\mathcal{E}_{\mathrm{film}} &= \frac{1}{2}\,
D_\mathrm{f}\, \bigl\langle \left( \Delta w \right)^2 \bigr\rangle
+  \frac{h}{2 E_\mathrm{f}} \bigl\langle \left( \Delta \phi \right)^2
\bigr\rangle
\\
\mbox{} &+ \frac{1}{2}\, \frac{h E_\mathrm{f}}{1 - \nu_\mathrm{H, f}^2} \left[ (\mathrm{tr}\, \mathbf{E})^2 - 2(1 - \nu_\mathrm{H, f}) \det \mathbf{E}\, \right] ,
\label{eqn:E_film}
\end{aligned}
\end{equation}
where $w(x, y)$ is the out-of-plane displacement of the film,
$\phi(x, y)$ is the Airy stress function that
obeys the equation
$\Delta^2 \phi = -E_\mathrm{f}\, [\partial_x^2 w\, \partial_y^2 w - (\partial_x \partial_y w)^2]$, and $\mathbf{E}$ is the mean strain tensor with components ${E}_{\alpha\beta} = -\bar{e} \delta_{\alpha\beta}
 + \frac{1}{2}\langle\partial_\alpha w\, \partial_\beta w\rangle$.
The angular brackets denote spatial averaging.
The other contribution to the total energy density is the film-substrate coupling:
\begin{equation}
\mathcal{E}_{\mathrm{sub}} = \frac{E^*}{4\Omega} \!\int\! \frac{d^2k}{(2\pi)^2} |\mathbf{k}|\, |\tilde{w}(\mathbf{k})|^2
- \frac{K_3}{3} {Q^2} \left\langle w^3(x, y) \right\rangle , 
\label{eqn:E_sub}
\end{equation}
where $\tilde{w}(\mathbf{k})$ is the Fourier transform of $w(x, y)$
and $\Omega$ is the total area. The first term in Eq.~\eqref{eqn:E_sub} is the standard result for a Hookean medium when the transverse components of the elastic deformation on its surface are neglected~\cite{Allan1969}. The second term is a correction that accounts for the asymmetry between compression and expansion in the normal direction. It is written assuming the dominant Fourier harmonics of the wrinkles have wavenumbers $|\mathbf{k}| \simeq Q = 2\pi/\lambda_\mathrm{w}$, as is the case near the critical point. We include this correction because such perturbations may influence the selection of wrinkle patterns that are in a close energetic competition~\cite{Huang2005, Cai2011, Brau2011}. 

We examined four competing states: the flat phase, the stripe phase,
the checkerboard phase, and the triangular phase~\cite{Chen2004, Huang2005, audolyboudaoud2008part1, Cai2011}.
Figure~\ref{fig:FvK_buckling_patterns}a shows the corresponding phase diagram.
For small wrinkle amplitude $A_\mathrm{w} = \max w(x,y)$, the three buckled states are well approximated by a superposition
$w(\mathbf{r}) = (A_\mathrm{w} / 2 n) \sum_j \exp \left(i \mathbf{Q}_j \cdot \mathbf{r}\right)$ of $2n$ plane waves ($n = 1$, $2$, or $3$)
whose wavevectors $\mathbf{Q}_j$ form a symmetric $2n$-point star~\cite{audolyboudaoud2008part1}.
These patterns are illustrated in Figs.~\ref{fig:FvK_buckling_patterns}b-d. For all the phases, the optimal period $\lambda_\mathrm{w}$
is very close to that predicted by Eq.~\eqref{eqn:lambda_w}. 

In agreement with previous work~\cite{Huang2005, audolyboudaoud2008part1, Cai2011}, we find that if $K_3 = 0$,
then the checkerboard pattern is favored above a certain threshold strain $e_\mathrm{c}$.
However, for a nonzero $K_3$, buckling into the triangular phase
occurs first.
The reason why a nonzero $K_3$ helps the triangular phase is because the spatial average of $w^3(x, y)$ is finite in the triangular phase but vanishes in the stripe and checkerboard phases.
As in other systems containing a cubic term in the Landau free-energy expansion, the buckling transition into the triangular phase is discontinuous,
preempting the second-order transition into the checkerboard phase.
For a given $K_3 \neq 0$, the triangular phase continues to dominate over the checkerboard phase over a range of $\bar{e}$.
Alternatively, at a fixed overstrain $\bar{e} - e_\mathrm{c} > 0$, the triangular phase wins within our model as long as $K_3$ exceeds  
$K_{3,\mathrm{c}}(\bar{e}) = c_2 E^* (\bar{e} -e_\mathrm{c})^{1/2} / e_\mathrm{c}$.
Here $c_2 = 0.038$ is a numerical coefficient derived in SI. For example, in our system $K_{3,\mathrm{c}} \sim 0.6\, \mathrm{GPa}$ at $10\%$ overstrain.
The value $K_3 = 1\, \mathrm{GPa}$ used in Figs.~\ref{fig:FvK_buckling_patterns}b-d 
was chosen based on the following reasoning.
We estimate $K_3 \sim 20\,\mathrm{GPa}$ if we use the third-order elastic constants of single-crystal graphite~\cite{Cousins2003} (SI).
However, these single-crystal values are expected to be reduced 
for pyrolytic graphite.
Assuming the same reduction factor of $5.0 / 0.26 \approx 20$ as for
$C_{44}$~\cite{Bosak2007, Blakslee1970}, we obtained $K_3 \sim 1\,\mathrm{GPa}$. 

In summary, we have observed that IX emission in a MoSe$_2$/WSe$_2$ van der Waals heterostructure exhibits a triangular spatial pattern with an interpeak distance of $\sim 2.6\,\mu\mathrm{m}$, which greatly exceeds the $10$-$\mathrm{nm}$ scale of typical moir\'e superlattices. This gradual modulation is distinct from short-range disorder.
It coexists with the long-range IX transport in the system.

The observed quasi-periodic pattern cannot be explained by the Turing or attractive dipole interaction instabilities. 
We attribute it instead to a structural transition: the wrinkling of the thin hBN film on the graphite substrate. Our modeling shows that the characteristic wavelength of the wrinkles should be in the micrometer range, and their preferred symmetry should be triangular, consistent with the observed emission pattern.

Since strain-induced wrinkling in layered systems with different elastic moduli is a general phenomenon, it could well be widespread in van der Waals heterostructures. Indeed, similar albeit less regular modulations have been observed in Refs.~\cite{Jauregui2019, Zhang2024}, with characteristic periods also in the $\mu\mathrm{m}$-range.
Future work will examine the effect of such buckling patterns on the electronic structure and collective phenomena in these two-dimensional materials.

\vskip5mm \textbf{Supporting Information.} SI presents additional descriptions of the heterostructure, experimental procedures, and the theoretical model.

\vskip 5mm \textbf{Acknowledgements.} We are grateful to A.~H.~MacDonald for discussions and to A.~K.~Geim for teaching us manufacturing of van der Waals heterostructures. The experiments were supported by the Department of Energy, Office of Basic Energy Sciences, under award DE-FG02-07ER46449. The heterostructure manufacturing was supported by NSF Grant 1905478. The data analysis was supported by NSF Grant 2516006. D.E.P. acknowledges support from NSF CAREER award no. DMR-2542485.


\end{document}


\date{\today}

\title{Supporting Information for
``Excitonic pattern formation from a wrinkling instability in a van der Waals heterostructure''}

\author{Zhiwen~Zhou}
\affiliation{Department of Physics, University of California San Diego, La Jolla, CA 92093, USA}
\author{L.~H.~Fowler-Gerace}
\affiliation{Department of Physics, University of California San Diego, La Jolla, CA 92093, USA}
\author{W.~J.~Brunner}
\affiliation{Department of Physics, University of California San Diego, La Jolla, CA 92093, USA}
\author{E.~A.~Szwed}
\affiliation{Department of Physics, University of California San Diego, La Jolla, CA 92093, USA}
\author{Michael~M.~Fogler}
\affiliation{Department of Physics, University of California San Diego, La Jolla, CA 92093, USA}
\author{Daniel~E.~Parker}
\affiliation{Department of Physics, University of California San Diego, La Jolla, CA 92093, USA}
\author{L.~V.~Butov} 
\affiliation{Department of Physics, University of California San Diego, La Jolla, CA 92093, USA}

\maketitle

\renewcommand*{\thefigure}{S\arabic{figure}}

\section{Heterostructure}

Manufacturing of the MoSe$_2$/WSe$_2$ heterostructure studied in this work (Figs.~1a,b in the main text) is described in Ref.~\cite{Fowler-Gerace2024}, where the same heterostructure is used for studies of IX transport. Figure~1b in the main text showing a microscope image and the layers of the heterostructure is similar to Fig.~S2b in Ref.~\cite{Fowler-Gerace2024} also showing a microscope image and the layers of the heterostructure. The long WSe$_2$ and MoSe$_2$ edges (Fig.~1b in the main text) enable a rotational alignment between the WSe$_2$ and MoSe$_2$ monolayers. The angle between the long MoSe$_2$ and WSe$_2$ edges in the heterostructure gives the twist angle $\delta \theta \sim 1.1^\circ$ as outlined in Ref.~\cite{Fowler-Gerace2024}. The IX $g$-factor $\sim - 16$ in the heterostructure corresponds to H stacking~\cite{Seyler2019, Wozniak2020}.

The accuracies of estimating $\delta \theta$ using the long WSe$_2$ and MoSe$_2$ edges and using SHG are comparable. We do not use SHG for additional $\delta \theta$ estimates because the intense excitation pulses in SHG measurements may cause a heterostructure deterioration and may suppress the efficient long-range IX transport~\cite{Fowler-Gerace2024, Zhou2024} in the heterostructure. To keep the quality of the sample, we also do not perform other measurements and limit the measurements by the optical measurements outlined in this work.

\begin{figure}[b]
\begin{center}
\includegraphics[width=5.5cm]{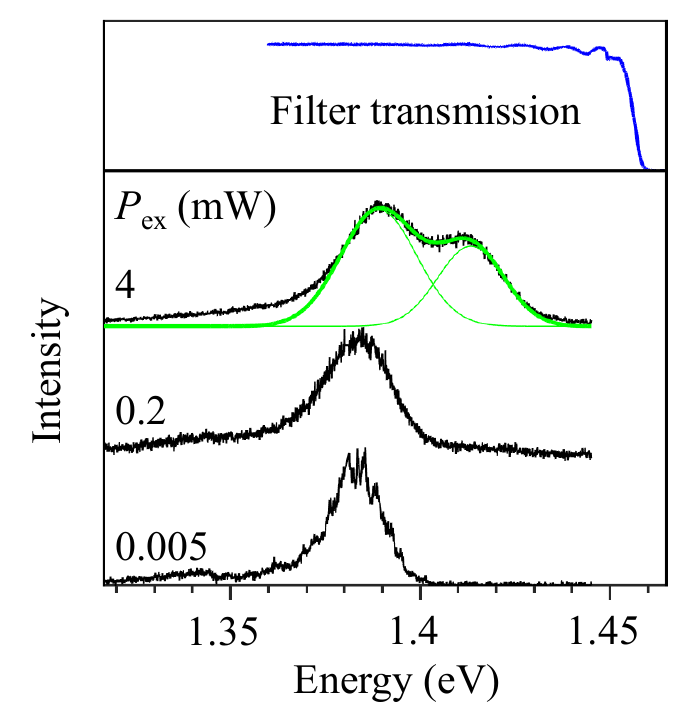}
\caption{IX PL filtering. IX PL spectra at different excitation powers $P_{\rm ex}$. $T = 1.7$~K. 
For $P_{\rm ex} = 4$~mW, the gaussian fits to the lower-energy IX line and higher-energy IX line are shown by the thin green lines, the sum of the gaussians is shown by the thick green line. The transmission spectrum of the filter used for imaging the IX PL patterns is shown on the top graph. The filter transmits the entire PL of IXs.
}
\end{center}
\label{fig:spectra}
\end{figure}

\begin{figure}[t]
\begin{center}
\includegraphics[width=8.5cm]{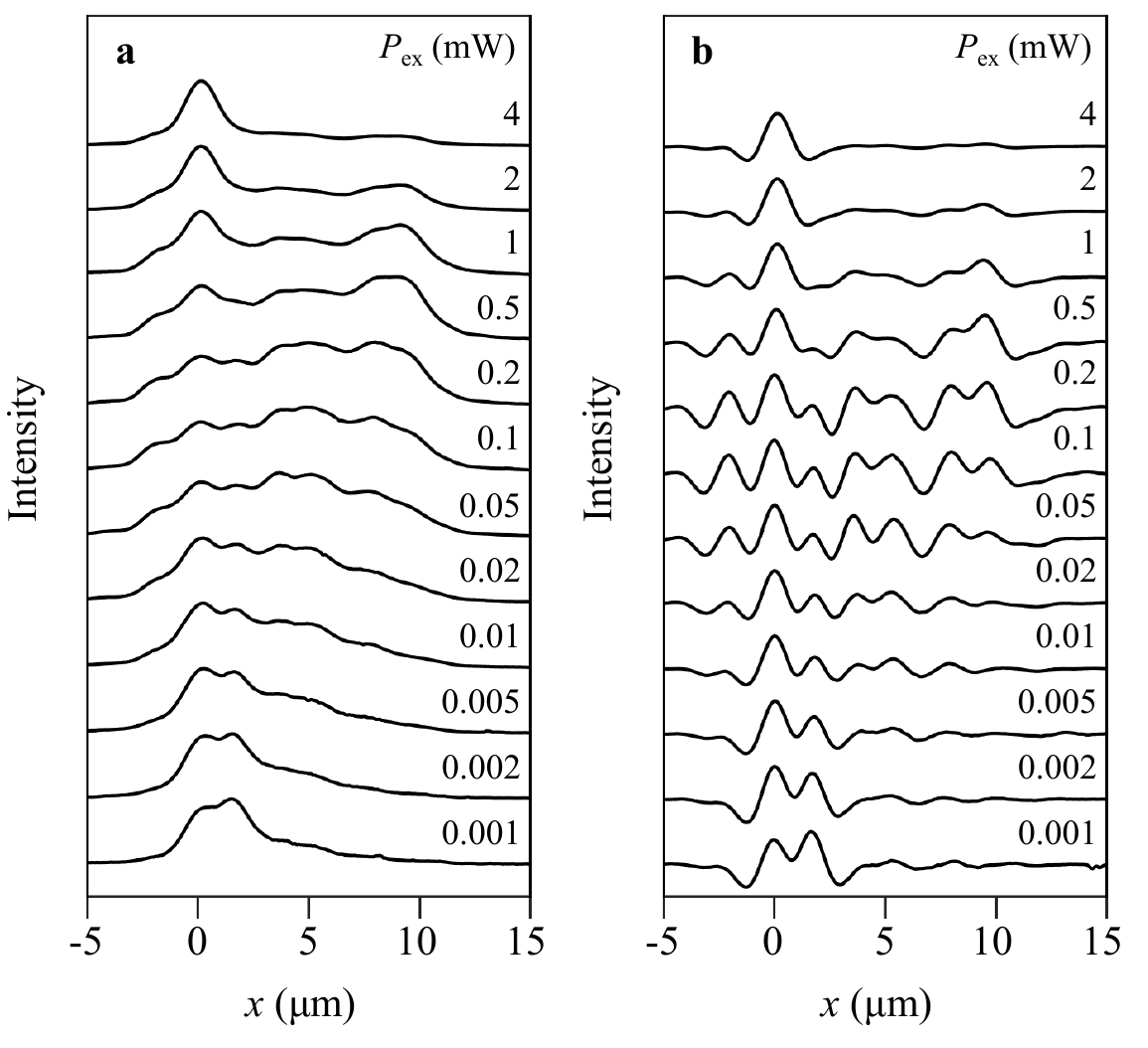}
\caption{Figure~S2 is similar to Fig.~3a,b in the main text, however, Fig.~3a,b shows $I(x)$ and $-\Delta I(x)$ for the lower-energy IX line, while Fig.~S2 shows $I(x)$ and $-\Delta I(x)$ for the IX PL given by the sum of lower-energy IX line and high-energy IX line.
}
\end{center}
\label{fig:spectra1}
\end{figure}

\begin{figure}[b]
\begin{center}
\includegraphics[width=9cm]{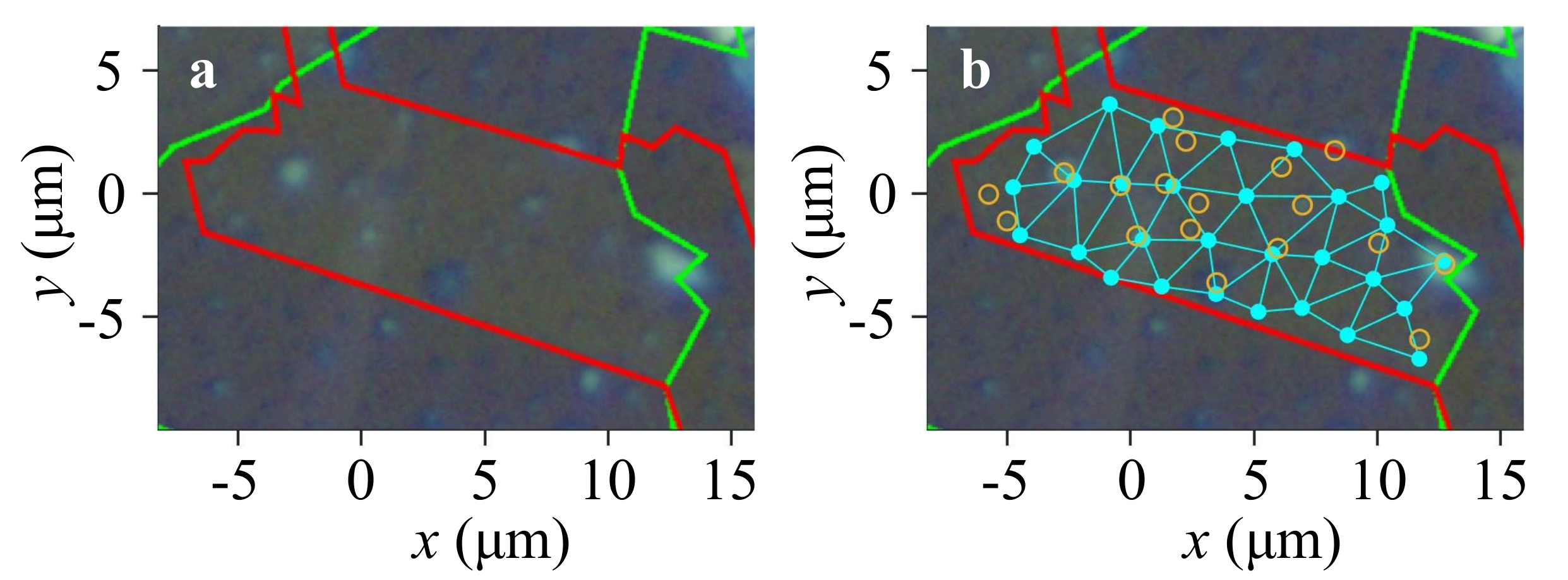}
\caption{Bubbles and local IX PL maxima. 
(a) A microscope image of the heterostructure (similar to Fig.~1b in the main text). 
The red and green lines indicate the boundaries of MoSe$_2$ and WSe$_2$ monolayers, respectively.
The positions of the bubbles in this image are shown by orange open circles in (b). 
(b) also shows the positions of the local IX PL maxima by cyan dots (same positions are shown in Figs.~1d,f in the main text). 
}
\end{center}
\label{fig:spectra2}
\end{figure}

\begin{figure*}[t]
\begin{center}
\includegraphics[width=12.5cm]{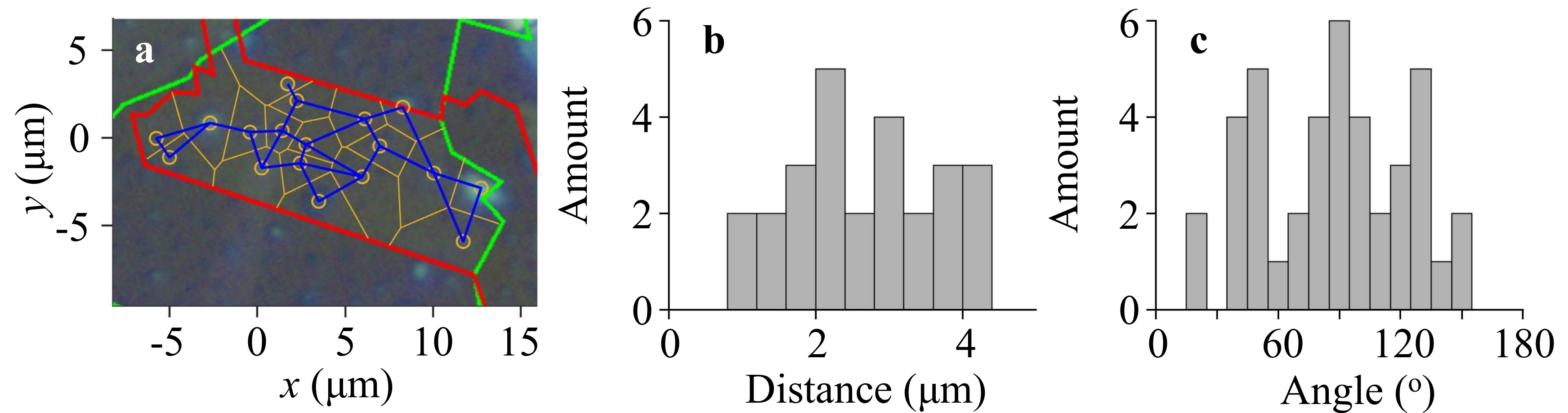}
\caption{Analysis of the bubble pattern. 
(a) The Voronoi diagram. The bubbles are shown by the orange circles. The
nearest neighbors are connected by the blue lines. The
Voronoi cells are marked by the orange lines. 
(b) The histogram of the nearest-neighbor distances. 
(c) The histogram of the angles between the nearest-neighbor lines. 
The large spread of both quantities indicates that the bubble positions are disordered. 
}
\end{center}
\label{fig:spectra3}
\end{figure*}

\section{IX PL filtering}

As outlined in the description of the optical measurements in the main text, the IX PL was selected by a filter $E < 1.46$~eV. Figure~S1 shows the IX PL spectra and the transmission spectrum of the filter used for imaging the IX PL patterns. The filter transmits the entire PL of IXs (Fig.~S1) so the images in Figs.~1 and 2 in the main text present the patterns of the entire IX PL.

A higher-energy IX PL line appearing at high $P_{\rm ex} > 0.2$~mW (Fig.~S1 and Fig.~3 in Ref.~\cite{Zhou2024}) was attributed to the appearance of moir{\'e} cells with double occupancy in Ref.~\cite{Zhou2024}. Figure~3a,b in the main text shows $I(x)$ and $-\Delta I(x)$ for the lower-energy IX line. For $P_{\rm ex} > 0.2$~mW in Fig.~3a,b in the main text, the lower-energy IX line is separated from the high-energy IX line by the gaussian fits as shown in Fig.~S1. Including the higher-energy IX line makes no significant effect on the intensity modulation: Fig.~S2 shows $I(x)$ and $-\Delta I(x)$ for the IX PL given by the sum of lower-energy IX line and high-energy IX line and the modulations in Fig.~S2 are similar to the modulations in Fig.~3a,b.

\section{Bubbles}

Figure~S3 provides comparison between the positions of the bubbles seen in the microscope images of the heterostructure and the positions of the local PL maxima in the IX triangular pattern in Figs.~1 and 2 in the main text. Figure~S3 shows that there is no significant correlation between the positions of the bubbles and those of the PL maxima. To quantify this, we performed an analysis of the pattern of bubbles (Fig.~S4) similar to the analysis of the local PL maxima in the IX pattern (Fig.~2). Compared to the IX pattern, the pattern of bubbles is much more disordered.
The bubbles have a broad variance of the coordination numbers of the Voronoi cells (Fig.~S4a), a broad variance of the nearest-neighbor distances (Fig.~S4b), and a broad variance of the angles between the lines connecting the neighboring bubbles (Fig.~S4c). Therefore, it is unlikely that the bubbles are responsible for the formation of the observed relatively well ordered IX pattern.

\section{Model}
\label{sec:Model}

\begin{figure}[b]
\begin{tikzpicture}[
    >=Latex,
    font=\itshape,
    every node/.style={font=\itshape}
]
\def\dx{1.8}
\def\dy{1.0}

\coordinate (FBL) at (0,0);
\coordinate (FBR) at (6,0);
\coordinate (FTL) at (0,3);
\coordinate (FTR) at (6,3);
\coordinate (BBL) at ({0+\dx},{0+\dy});
\coordinate (BBR) at ({6+\dx},{0+\dy});
\coordinate (BTL) at ({0+\dx},{3+\dy});
\coordinate (BTR) at ({6+\dx},{3+\dy});
\fill[gray!20] (FTL) -- (FTR) -- (BTR) -- (BTL) -- cycle;
\draw (FTL) -- (FTR) -- (BTR) -- (BTL) -- cycle;
\draw (FBR) -- (FTR) -- (BTR) -- (BBR) -- cycle;
\draw (FBL) -- (FBR) -- (FTR) -- (FTL) -- cycle;
\draw[dashed] (BTL) -- (BBL);
\draw[dashed] (BBL) -- (FBL);
\draw[dashed] (BBL) -- (BBR);
\def\layerT{0.28}
\coordinate (FTRl) at ($(FTR)-(0,\layerT)$);
\coordinate (BTRl) at ($(BTR)-(0,\layerT)$);
\fill[gray!40] (FTR) -- (BTR) -- (BTRl) -- (FTRl) -- cycle;
\draw (FTR) -- (BTR);
\draw (FTRl) -- (BTRl);
\coordinate (FTLl) at ($(FTL)-(0,\layerT)$);
\fill[gray!40] (FTL) -- (FTR) -- (FTRl) -- (FTLl) -- cycle;
\draw (FTLl) -- (FTRl);
\draw (FTL) -- (FTLl);
\draw (FTR) -- (FTRl);
\coordinate (BTLl) at ($(BTL)-(0,\layerT)$);
\draw[dashed] (BTLl) -- (FTLl);
\draw[dashed] (BTLl) -- (BTRl);

\coordinate (O) at (FTLl);
\draw[->] (O) -- ++(1.3,0) node[below, yshift=-3pt] {$x$};
\draw[->] (O) -- ++({0.55*\dx},{0.55*\dy}) node[right, xshift=4pt, yshift=-3pt] {$y$};
\draw[->] (O) -- ++(0,1.3) node[above] {$z$};
\fill (O) circle (1pt);
\coordinate (Q) at (3.6,2.15);
\node (ulabel) at (Q) {$\mathbf{u} = (u,v,w)$};
\def\fh{0}
\coordinate (RefTop) at ($(FTR)+({\fh*\dx},{\fh*\dy})$);
\coordinate (RefBot) at ($(RefTop)-(0,\layerT)$);
\coordinate (RefTopExt) at ($(RefTop)+(0.75,0)$);
\coordinate (RefBotExt) at ($(RefBot)+(0.75,0)$);
\draw (RefTop) -- (RefTopExt);
\draw (RefBot) -- (RefBotExt);
\draw[->] ($(RefTopExt)+(0,0.55)$) -- (RefTopExt);
\draw[->] ($(RefBotExt)+(0,-0.55)$) -- (RefBotExt);
\node[right] at ($(RefTopExt)!0.5!(RefBotExt)+(0.05,0)$) {$h$};
\coordinate (TopFaceY) at (0,3.3);
\node[font=\normalfont] at (TopFaceY -| ulabel) {$E_\mathrm{f}$};
\coordinate (BotFaceY) at (0,1.5);
\node[font=\normalfont] at (BotFaceY -| ulabel) {$E^*$};
\end{tikzpicture}
\caption{Geometry of the problem.
}
\label{fig:schematics}
\end{figure}

Consider a thin film of thickness $h$ resting on a semi-infinite elastic substrate occupying the half-space $z < 0$,
see Fig.~\ref{fig:schematics}.
The film and substrate are bonded, so
that the vector of elastic deformations $\mathbf{u} = (u, v, w)$
is continuous in space.
We use $\alpha,\beta \in \{ x,y\}$ for in-plane indices
(e.g., $u = u_x$, $v = u_y$)
and $i, j, k \in \{x, y, z\}$ for 3D indices, e.g., $u_z = w$.
Repeated indices are summed over.
For example, the 2D Laplacian operator can be written as $\Delta \equiv \partial_\alpha \partial_\alpha = \partial_x^2 + \partial_y^2$.
The energy of the film in our model is calculated according to the F\"oppl-von K\'arm\'an (FvK) theory of thin elastic plates~\cite{landau2012theory}. To make the presentation more self-contained,
we begin with summarizing the key elements of this theory.

The full form of the elastic strain tensor is
\begin{equation}
\label{eqn:strain_def}
	e_{i k}(x, y, z) = -\bar{e}_{i k} + \frac{1}{2}\left( \partial_i u_k + \partial_k u_i \right)
	+ \frac{1}{2}  
	\partial_i u_j \partial_j u_j\,,
\end{equation}
where, compared to the standard definition~\cite{landau2012theory},
we added $-\bar{e}_{i k}$, the strain in the pre-buckled state.
We take this pre-strain to exist only in the film and to be uniform and isotropic,
\begin{equation}
	\bar{e}_{x x} = \bar{e}_{y y} = \bar{e} = \text{const}\,,
    \qquad 0 < z < h\,,
\label{eqn:bar_e}    
\end{equation}
with $\bar{e} > 0$ and all other $\bar{e}_{i k}$ being equal to zero everywhere. 
The right-hand side of Eq.~\eqref{eqn:strain_def} contains terms quadratic in $u_j$. In the FvK theory, among such terms, only those quadratic in $w$ are retained~\cite{landau2012theory}.
Hence, the in-plane strain components become
\begin{equation}
e_{\alpha \beta}(x,y) = -\bar{e}_{\alpha \beta}
    + \frac{1}{2}\left( \partial_\alpha u_\beta + \partial_\beta u_\alpha \right)
	+ \frac{1}{2}\,  
	\partial_\alpha w\, \partial_\beta w
\label{eqn:strain_FvK}    
\end{equation}
inside the film, $0 < z < h$.
Here $u$, $v$, and $w$ are to be evaluated
at the film's midplane $z = \frac{h}{2}$, so that each $e_{\alpha\beta}(x,y)$ is a function of $x$ and $y$ only.
If the film is flat, $w(x, y) = 0$, then the average value of $e_{\alpha \beta}$
is equal to $-\overline{e}_{\alpha \beta}$ because
the averages of full derivatives vanish, e.g.,
\begin{equation}
\braket{\partial_\alpha u_\beta} = \frac{1}{\Omega}
\int \partial_\alpha u_\beta\, d x d y = 0\,,
\end{equation}
where $\Omega$ is the system area and we assume periodic boundary conditions in the $x$--$y$ plane.
If the film is bent, then the average value of $e_{\alpha\beta}$ becomes
\begin{equation}
	{E}_{\alpha\beta} \equiv \braket{e_{\alpha\beta}}
    = -\overline{e}_{\alpha\beta}
    + \frac{1}{2}\braket{\partial_\alpha w\, \partial_\beta w} .
\label{eqn:average_strain}    
\end{equation}
This is a key relation behind the buckling instability:
the buckled film has a smaller (by absolute value) average strain,
and so it can have a lower total energy if the balance between stretching and bending energy is optimized.

To calculate these various energy components we need expressions for
the elastic stress.
We assume that the elastic media under consideration --- the film and the substrate --- are hexagonal crystals, in which case they are characterized by
five independent elastic constants $C_{11}$, $C_{12}$, $C_{13}$, $C_{33}$, and $C_{44}$, same as
in the case of a full transverse isotropy.
Table~\ref{tb:elastic_constants} lists these constants
for graphite and hBN.

\begin{table*}
\begin{tabular}{l S S S S[table-format=3.4] S[table-format=2.4]%
S[table-format=4.0] S[table-format=4.0] S S S S}
\hline\\
{Material}
  & {$C_{11}$} & {$C_{12}$} & {$C_{13}$} & {$C_{33}$}
  & {$C_{44}$}
  & {$E$}   & {$\overline{E}$}  & {$E^*$} & $M$ & {2$\sqrt{C_{33} C_{44}}$}\\[3pt]
\hline\\
{Crystalline graphite~\cite{Bosak2007}}
  & 1109(16) & 139(36) & 0(3)  & 38.7(7) & 5.0(3)
  & 1092 & 1109 & 27.8 & 27.2 & 27.8\\
{Pyrolytic graphite~\cite{Blakslee1970}}
  & 1060(20) & 180(20) & 15(5) & 36.5(10) & 0.26(8)
  & 1025 & 1054 & 6.14 & 6.14 & 6.16\\
{hBN}~\cite{Bosak2006}
  & 811(12)  & 169(24) & 0(3) & 27.0(5) & 7.7(5)
  & 776  & 811 & 28.8 & 27.4 & 28.8\\[3pt]
\hline
\end{tabular}
\caption{Elastic constants $C_{J K}$ of graphite and hBN (in GPa)
and the corresponding elastic moduli $E$, $\overline{E} = E / (1 - \nu_\mathrm{H}^2)$, $E^*$, and $M$ computed according to Eqs.~\eqref{eqn:nu}, \eqref{eqn:E*}, and \eqref{eqn:M}.
The rightmost column [Eq.~(1) of the main text] is an approximation for
both $E^*$ and $M$.
We use the parameters of Refs.~\cite{Blakslee1970} and \cite{Bosak2006} for the film and the substrate, respectively, in our calculations.
}
\label{tb:elastic_constants}
\end{table*}

The relation between the elements of the 3D stress tensor $\sigma_J$ and those of the 3D engineering strain $\eta_K$ ($J, K = 1, 2, \ldots 6$) of such media is as follows:
\begin{equation}
\begin{pmatrix}
\sigma_{xx} \\
\sigma_{yy} \\
\sigma_{zz} \\
\sigma_{xz} \\
\sigma_{yz} \\
\sigma_{xy}
\end{pmatrix}
=
\begin{pmatrix}
C_{11} & C_{12} & C_{13} & 0      & 0      & 0                          \\
C_{12} & C_{11} & C_{13} & 0      & 0      & 0                          \\
C_{13} & C_{13} & C_{33} & 0      & 0      & 0                          \\
0      & 0      & 0      & C_{44} & 0      & 0                          \\
0      & 0      & 0      & 0      & C_{44} & 0                          \\
0      & 0      & 0      & 0      & 0      & \tfrac{1}{2}(C_{11}-C_{12})
\end{pmatrix}
\begin{pmatrix}
e_{xx} \\
e_{yy} \\
e_{zz} \\
2e_{xz} \\
2e_{yz} \\
2e_{xy}
\end{pmatrix}
\,,
\label{eqn:sigma_from_e}
\end{equation}
where $e_{x x} = \eta_1$, $e_{y y} = \eta_2$, $\dots$, and $2 e_{x y} = \eta_6$.
The matrix above is known as the stiffness matrix.
Its inverse is the compliance matrix:
\begin{equation}
\frac{1}{E}\
\begin{pmatrix}
1      & -\nu_H & -\nu_V & 0        & 0        & 0           \\
-\nu_H & 1      & -\nu_V & 0        & 0        & 0           \\
-\nu_V & -\nu_V & \lambda & 0        & 0        & 0           \\
0      & 0      & 0      & 2(1+\nu_4) & 0        & 0           \\
0      & 0      & 0      & 0        & 2(1+\nu_4) & 0           \\
0      & 0      & 0      & 0        & 0        & 2(1+\nu_H)
\end{pmatrix}\,,
\label{eqn:e_from_sigma}
\end{equation}
where the elastic modulus $E$ and the four dimensionless coefficients $\nu_H$, $\nu_V$, $\nu_4$, and $\lambda$ are given by
\begin{gather}
\nu_H = \frac{C_{12} C_{33} - C_{13}^2}{C_{31}^2 - C_{13}^2} ,
\quad
\nu_V = C_{13}\, \frac{C_{11} - C_{12}}{C_{31}^2 - C_{13}^2} ,
\\
\lambda = \frac{C_{11}^2 - C_{12}^2}{C_{31}^2 - C_{13}^2} ,
\quad
\nu_4 = \frac12 \frac{E_s}{C_{44}} - 1\,,
\label{eqn:lambda}\\
E = C_{11} - \nu_H C_{12} - \nu_V C_{13} \,,
\label{eqn:nu}
\end{gather}
and we used another convenient notation~\cite{Gross2013}
\begin{equation}
C_{31} \equiv \sqrt{C_{11} C_{33}}\,.
\label{eqn:C_31}
\end{equation}

In the FvK theory,
the following strains components are assumed to be negligible:
\begin{equation}
\sigma_{x z} = \sigma_{y z} = \sigma_{z z} = 0\,,
\qquad 0 < z < h\,.
\label{eqn:sigma_xz_FvK}
\end{equation}
Therefore,
\begin{equation}
e_{x z} = e_{y z} = 0\,,
\qquad 0 < z < h\,,
\label{eqn:e_xz_FvK}
\end{equation}
while the relations between $\sigma_{\alpha\beta}(x,y)$ and $e_{\alpha\beta}$ simplify to
\begin{align}
\sigma_{xx} &= \frac{E_\mathrm{f}}{1-\nu^2} \left( e_{xx} + \nu e_{yy} \right),
\\
\sigma_{yy} &= \frac{E_\mathrm{f}}{1-\nu^2} \left( e_{yy} + \nu e_{xx} \right),
\\
\sigma_{xy} &= \frac{E_\mathrm{f}}{1 + \nu}\, e_{xy} \,,
\label{eqn:sigma_from_e_FvK}    
\end{align}
where $E_\mathrm{f}$ is the Young modulus of the film
and we use $\nu$ instead of $\nu_\mathrm{H, f}$ in order to lighten the notations.
Omitting further details of the derivation~\cite{landau2012theory},
we write down the final result for the FvK energy functional of the film:
\begin{align}
	E_{\mathrm{film}} &= E_{\mathrm{bend}} + E_{\text{stretch}}\,,
\label{eqn:E_FvK1}\\    
	E_{\mathrm{bend}}[w] &= \frac{D_\mathrm{f}}{2}\int \left\{ \left( \Delta w \right)^2 - (1-\nu) {[w,w]} \right\}
    d x d y ,
\label{eqn:E_bend1}\\
	E_{\text{stretch}} &= \frac{h}{2} \int \sigma_{\alpha\beta} e_{\alpha\beta}\,
    d x d y\,.
\label{eqn:E_stretch1}
\end{align}
Here $D_\mathrm{f} = \overline{E}_f h^3 / 12$ is the bending rigidity and
$[w,w]$ stands for $2\partial_x^2 w\, \partial_y^2 w - 2(\partial_x \partial_y w)^2$.
In general, the Monge-Amp\'ere bracket $[f, g]$ of two functions $f(x, y)$ and $g(x, y)$ is defined by
\begin{equation}
[f, g] = 
(\partial_x^2 f)(\partial_y^2 g)
+ (\partial_y^2 f) (\partial_x^2 g)
- 2 (\partial_x \partial_y f) (\partial_x \partial_y g) .
\end{equation}
It is a total divergence, so it vanishes upon integration under the assumed periodic boundary conditions.

Taking the variational derivative of $\mathcal{E}_{\text{stretch}}$ with respect to $u_\alpha$ yields the force balance equation
\begin{equation}
	\partial_\beta \sigma_{\alpha\beta} = 0\,.
\label{eqn:force_balance1}    
\end{equation}
Its solution can be represented as
\begin{equation}
	\sigma_{\alpha\beta}(x,y) = \braket{\sigma_{\alpha\beta}}
    + \hat{\sigma}_{\alpha\beta}(x, y)\,,
\end{equation}
where the first term is a constant (the average strain)
and the second term $\hat{\sigma}_{\alpha\beta}(x, y)$
is a function of zero mean
that can be expressed in terms of some scalar function $\phi(x,y)$
known as the Airy stress function:
\begin{equation}
	\label{eqn:Airy_function}
	\hat{\sigma}_{xx} = \partial_y^2 \phi\,,
    \qquad
    \hat{\sigma}_{yy} = \partial_x^2 \phi\,,
    \qquad
    \hat{\sigma}_{xy} = -\partial_x\partial_y \phi\,.
\end{equation}
From Eqs.~\eqref{eqn:strain_FvK}, \eqref{eqn:sigma_from_e_FvK}, and \eqref{eqn:Airy_function}
one can show that $\phi(x,y)$ obeys the equation
\begin{equation}
	\Delta^2 \phi = -\frac{1}{2}\, E_\mathrm{f}\, [w,w]\,,
\label{eqn:Airy}
\end{equation}
which is referred to as the compatibility condition.
Rewriting the energy functionals in terms of $w$ and $\phi$,
we arrive at the formula for the energy density of the film
\begin{equation}
\begin{split}
\mathcal{E}_{\mathrm{film}} &= \frac{E_{\mathrm{film}}}{\Omega}
= \frac{1}{2} \int 
    \left[D_\mathrm{f} \left( \Delta w \right)^2 + 
\frac{h}{E_\mathrm{f}} \left( \Delta \phi \right)^2 \right] \frac{d x d y}{\Omega}
\\
\mbox{} &+ \frac{1}{2}\, \frac{h E_\mathrm{f}}{1 - \nu^2} \left[ (\mathrm{tr}\, \mathbf{E})^2 - 2(1 - \nu) \det \mathbf{E}\, \right] ,
\end{split}
\label{eqn:E_film}
\end{equation}
where we have taken into account that $[w, w]$ and $[\phi, \phi]$ drop out upon integration. This is Eq.~(3) of the main text.

The total energy density of the system is
\begin{equation}
\mathcal{E} = \mathcal{E}_{\mathrm{film}} + \mathcal{E}_{\text{sub}}\,,
\label{eqn:E_FvK}
\end{equation}
where the film-substrate interaction term $\mathcal{E}_{\mathrm{sub}}$
will be established in the next section, Sec.~\ref{sec:substrate}.

Before we go there, we would like to recount an interesting connection~\cite{audoly2010elasticity} to Gauss's \textit{Theorema Egregium} that states, roughly, that the extrinsic Gaussian curvature of a surface is equal to its intrinsic curvature.
To see this relation, we treat the film as a Riemann manifold
whose metric tensor has components $g_{\alpha\beta} = \delta_{\alpha\beta} + 2 e_{\alpha\beta}$.
The intrinsic curvature of such a manifold is equal to $-\inc \mathbf{e}$,
where $\inc \mathbf{A}$ is the incompatibility operator acting on an arbitrary rank-$2$ tensor $\mathbf{A}$ as follows:
\begin{equation}
	\inc \mathbf{A} = \partial^2_{y} A_{xx} + \partial^2_{x} A_{yy} - 2 \partial_x \partial_y A_{xy} .
\end{equation}
Application of this operator to the strain tensor yields
\begin{equation}
	\inc \mathbf{e} = -\frac{1}{2}\, [w,w]\,.
\end{equation}
On the other hand, the (extrinsic) Gaussian curvature of a surface $z = w(x,y)$ is
\begin{equation}
\begin{split}
K_G &= 
\frac{(\partial_x^2 w) (\partial_y^2 w) - (\partial_x \partial_y w)^2}
     {\left[ 1 + (\partial_ x w)^2 + (\partial_y w)^2 \right]^2}
\\     
\mbox{} &\approx \partial_x^2 w\, \partial_y^2 w - (\partial_x \partial_y w)^2
= \frac{1}{2}\, [w,w]\,.
\end{split}
\end{equation}
Therefore, the extrinsic curvature is equal to the intrinsic one.
Physically, this is because the out-of-plane displacements in the equilibrium state
must be linked to the in-plane stretches if the surface tractions are negligible, Eq.~\eqref{eqn:sigma_xz_FvK}.

\section{Elastic response of a semi-infinite substrate}
\label{sec:substrate}
\subsection{Linear response}

In this subsection we derive the formula for the surface stiffness matrix~\cite{Cai2011} of a semi-infinite medium occupying the domain $z < 0$, see Fig.~\ref{fig:schematics}.
This $3 \times 3$ matrix $\mathbf{B}_s(\mathbf{k})$
relates the vector of tractions $\mathbf{f} = (\sigma_{xz}, \sigma_{yz}, \sigma_{zz})\,|_{\,z = 0}$
to the vector of elastic deformations $\mathbf{u}\,|_{\,z = 0} = (u_0, v_0, w_0)$ at the surface of the medium. In terms of the Fourier amplitudes,
$\tilde{\mathbf{f}}(\mathbf{k}) = \int \mathbf{f}(x, y) e^{-i k_x x - i k_y y} d x d y$ and $\tilde{\mathbf{u}}(\mathbf{k}, z) = \int \mathbf{u}(x, y, z) e^{-i k_x x - i k_y y} d x d y$, the relation is
\begin{equation}
\tilde{\mathbf{f}}(\mathbf{k}) = \mathbf{B}_s(\mathbf{k}) \cdot \tilde{\mathbf{u}}(\mathbf{k}, 0)\,.
\label{eqn:B_s_def}
\end{equation}
As discussed originally by Allen~\cite{Allan1969},
the most important part of the surface stiffness matrix in the context of the wrinkling transition is $\left[\mathbf{B}_s(\mathbf{k})\right]_{33}$.
It determines the elastic modulus
\begin{equation}
E^* = \frac{2}{k} \left[\mathbf{B}_s(\mathbf{k})\right]_{33} ,
\qquad k = |\mathbf{k}| = \sqrt{k_x^2 + k_y^2} \,,
\label{eqn:E^*_def1}
\end{equation}
which is the proportionality constant between the
amplitudes  $\tilde{f}_{z}$ and $\tilde{w}_0$ of the normal force and the
normal deformation under the boundary condition that the transverse deformation is absent:
$\tilde{u}_0 = \tilde{v}_0 = 0$.
A related but different quantity is the indentation modulus $M$, given by
\begin{equation}
M = \frac{2}{k}\, \frac{1}{\left[\mathbf{B}_s^{-1}(\mathbf{k})\right]_{33}} \,,
\label{eqn:M_def}
\end{equation}
which is the ratio of $\tilde{f}_z$ and $\tilde{w}_0$ under the boundary condition that the transverse force is absent:
$\tilde{f}_{x} = \tilde{f}_{y} = 0$. This modulus is relevant for a frictionless indenter problem~\cite{Gross2013}.
In theory, the distinction between $E^*$ and $M$ is conceptually important~\cite{audolyboudaoud2008part1};
in practice, the difference is small~\cite{Cai2011}.
We will compare $E^*$ and $M$ below once we have derived the full matrix $\mathbf{B}_s$.

Following \S5.12 of \cite{Green1968}, it is easy to compute the Fourier amplitude $\tilde{\mathbf{u}}(\mathbf{k}, z)$ of the elastic deformation inside the medium given the boundary condition $\tilde{\mathbf{u}}(\mathbf{k}, 0) = (\tilde{u}_0, \tilde{v}_0, \tilde{w}_0)$ at the surface.
The solution is represented by the sum of three exponential functions,
\begin{equation}
\tilde{\mathbf{u}}(\mathbf{k}, z) = \sum_{i = 1}^3 \tilde{\mathbf{u}}_i(\mathbf{k}) \exp\left(\frac{k}{\mu_i} z\right) ,
\qquad
\tilde{\mathbf{u}}_i(\mathbf{k}) = \mathbf{U}_i(\mathbf{k}) \cdot \tilde{\mathbf{u}}(\mathbf{k}, 0)\,,
\label{eqn:u_decay}
\end{equation}
with the decay rates determined by dimensionless constants $\mu_i$ such that
\begin{align}
\frac{1}{\mu_1} &= \sqrt{\frac{\beta + \alpha}{2}} + \sqrt{\frac{\beta - \alpha}{2}} 
\approx \sqrt{\frac{C_{11}}{C_{44}}}
= 64 ,
\label{eqn:mu_1}\\
\frac{1}{\mu_2} &= \sqrt{\frac{\beta + \alpha}{2}} - \sqrt{\frac{\beta - \alpha}{2}}
\approx \sqrt{\frac{C_{44}}{C_{33}}}
= 0.085 ,
\label{eqn:mu_2}\\
\alpha &= \sqrt{\frac{C_{11}}{C_{33}}}
= 5.4 ,
\label{eqn:alpha}\\
\beta &= 
\frac{C_{31}^2 - C_{13}^2 - 2 C_{13} C_{44}}{2 C_{33} C_{44}}
\approx \frac{C_{11}}{2 C_{44}} = 2030 ,
\label{eqn:beta}\\
\frac{1}{\mu_3} &= \sqrt{\frac{C_{11} - C_{12}}{2 C_{44}}}
= 41
\label{eqn:mu_3}
\end{align}
(the numerical values are for graphite, estimated per Ref.~\cite{Blakslee1970},
see Table~\ref{tb:elastic_constants}).
The $3 \times 3$ matrices $\mathbf{U}_i$ in Eq.~\eqref{eqn:u_decay} are
\begin{align}
 \mathbf{U}_1 &= \frac{1}{r_1 - r_2}
\begin{pmatrix}
i n_x\\
i n_y\\
r_1
\end{pmatrix} 
\begin{pmatrix}
i r_2 n_x & i r_2 n_y & 1
\end{pmatrix} ,
\quad n_\alpha \equiv \frac{k_\alpha}{k} \,,
\\[6pt]
\mathbf{U}_2 &= \left.\mathbf{U}_1\right|_{r_1 \leftrightarrow r_2} ,
\\[6pt]
\mathbf{U}_3 &=
\begin{pmatrix}
n_y\\
-n_x\\
0
\end{pmatrix} 
\begin{pmatrix}
n_y & -n_x & 0
\end{pmatrix} .
\label{eqn:U}
\end{align}
They play the role of orthogonal projectors that satisfy the equations $\mathbf{U}_i \cdot \mathbf{U}_j = \delta_{ij}\, \mathbf{U}_i$ and
the completeness relation $\mathbf{U}_1 + \mathbf{U}_2 + \mathbf{U}_3 = \mathbf{1}$.
The notations in Eq.~\eqref{eqn:U} are as follows (cf.~\cite{Green1968}):
\begin{align}
r_1 &= \frac{k_1}{\mu_1} ,
\quad
r_2  = \frac{\alpha}{r_1} = \frac{k_2}{\mu_2} ,
\\
k_1 &= \frac{1}{k_2} = \frac{C_{13} + C_{44}}{C_{33} \mu_1^2 - C_{44}} .
\label{eqn:k_1}
\end{align}
Inserting $\tilde{\mathbf{u}}(\mathbf{k}, z)$ into Eq.~\eqref{eqn:sigma_from_e}
(or, more expediently, into equations in \S5.12 of \cite{Green1968}),
we obtain $E^*$ and $\mathbf{B}_s$:
\begin{align}
E^* &= 2\sqrt{C_{33} C_{44}}\,
\left[1 - \frac{(C_{13} + C_{44})^2}{\left(C_{31} + C_{44}\right)^2}\right]^{1/2}
\notag\\
\mbox{} &= 6.143\,\mathrm{GPa} \,,
\label{eqn:E*}\\[6pt]
\mathbf{B}_s &= \frac12\, E^* k\ 
\begin{pmatrix}
\alpha - \alpha\zeta n_y^2  & \alpha\zeta n_x n_y       & i \gamma n_x\\
\alpha\zeta n_x n_y     & \alpha - \alpha\zeta n_x^2    & i \gamma n_y\\
-i \gamma n_x      & -i \gamma n_y        & 1
\end{pmatrix} ,
\label{eqn:B_s}\\[6pt]
\gamma &= \sqrt{\frac{C_{44}}{C_{33}}} \left(\frac{C_{31} - C_{13}}{C_{31} + C_{13} + 2 C_{44}}\right)^{1/2}
\notag\\
\mbox{} &\approx \sqrt{\frac{C_{44}}{C_{33}}}
= 0.078,
\label{eqn:gamma}\\
\zeta &= 1 - \sqrt{\frac{C_{11} - C_{12}}{2 C_{11}}}\, \frac{2\sqrt{C_{33} C_{44}}}{E^*} = 0.35\,.
\label{eqn:zeta}
\end{align}
Computing the matrix inverse of $\mathbf{B}_s$, we find the indentation modulus~\cite{Gross2013}
\begin{equation}
\begin{split}
M &= 2\sqrt{C_{33} C_{44}}\, \left(1 + \frac{C_{13}}{C_{31}}\right)
\left(\frac{C_{31} - C_{13}}{C_{31} + C_{13} + 2 C_{44}}\right)^{1/2}
\\
\mbox{} &= 6.136\,\mathrm{GPa}\,.
\end{split}
\label{eqn:M}
\end{equation}
It is only slightly smaller than $E^*$.

The three decaying exponentials in Eq.~\eqref{eqn:u_decay} can be thought of as the evanescent zero-frequency sound waves due to, respectively, the quasi-longitudinal, the quasi-shear, and the shear-horizontal modes launched at the surface of the medium.
Due to the strong anisotropy, the quasi-shear mode described by $\mathbf{U}_2$ has by far the slowest decay rate $k / \mu_2$, i.e., the longest penetration depth, of the three. 
Furthermore, for graphite, $r_1 \sim 10^{-2}$ is small and $r_2 \sim 10^3$ is large. Neglecting terms of order $r_1$ and $1 / r_2$, we obtain
\begin{align}
 \mathbf{U}_1 &=
\begin{pmatrix}
n_x^2   & n_x n_y   & 0\\
n_x n_y & n_y^2     & 0\\
0       &           & 0
\end{pmatrix} ,
\\[6pt]
\mathbf{U}_2 &=
\begin{pmatrix}
0 & 0 & 0\\
0 & 0 & 0\\
0 & 0 & 1
\end{pmatrix} ,
\\[6pt]
\mathbf{U}_3 &=
\begin{pmatrix}
 n_y^2    & -n_x n_y & 0\\
-n_x n_y  &  n_x^2   & 0\\
0         & 0        & 0
\end{pmatrix} ,
\label{eqn:U_simple}
\end{align}
which implies a simple single-exponential decay of $\tilde{w}$ with $z$:
\begin{align}
\tilde{w}(\mathbf{k}, z) = \tilde{w}_0(\mathbf{k}) \exp\left(\frac{k}{\mu_2} z\right) .
\label{eqn:w_z}
\end{align}
If the surface deformation contains multiple Fourier harmonics with different
wavevectors $\mathbf{k}$,
the response can be computed by summing their individual contributions.
In the equilibrium state near the critical point for the wrinkling instability, all $|\mathbf{k}| \simeq k_c$ are nearly the same.
As mentioned in the main text, a good approximation of wrinkling patterns we studied is
a linear combination of plane waves 
whose wavevectors have an equal length $Q$ and form a
symmetric $2n$-point star,
\begin{equation}
w_0(x, y) = \frac{1}{n}\, A_\mathrm{w}
\sum_{j = 0}^{n - 1} \cos \left[Q \left(\cos \frac{\pi}{n} j\right) x +  Q \left(\sin \frac{\pi}{n} j\right) y - \theta\right] .
\label{eqn:w0_triangular}
\end{equation}
This is similar to \cite{audolyboudaoud2008part1, Cai2011} but with an
extra phase shift $\theta$. With this addition, we can assume that the overall amplitude $A_\mathrm{w}$ of the wrinkles is positive without loss of generality.
For such $w_0(x, y)$, the vertical deformation $w$ at an arbitrary depth inside of the substrate has the same lateral profile as it does on the surface:
\begin{equation}
w(x, y, z) = w_0(x, y) \exp\left(-\sqrt{\frac{C_{44}}{C_{33}}}\, Q\, |z|\right) .
\label{eqn:w_z_critical}
\end{equation}

\subsection{Nonlinear correction}

In this subsection we estimate corrections to the substrate energy due to elastic nonlinearities. We will consider the terms up to the third power in deformations.
In order to collect all such terms we need to include the second and the third order terms in the expansion of the substrate energy density in powers of strain:
\begin{equation}
\frac{1}{2} \sum_{I, J}^6 C_{I J}\eta_I \eta_J +
\frac{1}{6} \sum_{I, J, K = 1}^6 C_{I J K}\eta_I \eta_J \eta_K + \ldots
\label{eqn:E_sub_expansion}
\end{equation}
Here $C_{I J}$ are the usual elastic constants (see Sec.~\ref{sec:Model}) and $C_{I J K}$ are Brugger's third-order constants.
Although the first sum is only quadratic in strain,
the strains themselves include quadratic corrections, see Eq.~\eqref{eqn:strain_def}. As a result, the product $\eta_1 \eta_3 = e_{x x} e_{z z}$
contains $\bigl(\partial_x w\bigr)^2 \partial_z w$, which is cubic in $w$. The same term is contained in $\eta_3 \eta_4^2 = 4 e_{z z} e_{x z}^2$.
The nonlinearities that come from $\eta_I \eta_J$ have a geometric origin and those that come from $\eta_I \eta_J \eta_K$ are due to material response. Both need to be accounted for.

Near the critical point, the in-plane deformations $u$ and $v$ are of order $O(Q A_\mathrm{w}^2)$.
They are much smaller than the out-of-plane deformation $w = O(A_\mathrm{w})$.
The strains $\eta_1 = e_{xx}$, $\eta_2 = e_{yy}$, and $\eta_6 = 2 e_{xy}$ are of order $O(Q^2 A_\mathrm{w}^2)$ and
are much smaller than $\eta_3 \simeq \partial_z w$, $\eta_4 \simeq \partial_y w$, and $\eta_5 \simeq \partial_x w$, which are of order $O(Q A_\mathrm{w})$.
Collecting all the dominant terms,
we get the total cubic correction to the energy
\begin{widetext}
\begin{align}
E^{(3)}_\mathrm{sub} &= \int \left\{\left(\frac{1}{2}\, C_{33} + \frac{1}{6}\, C_{333} \right)
\bigl(\partial_z w\bigr)^3
+ \left( \frac{1}{2}\, C_{13} + C_{44} + \frac{1}{2}\, C_{344}\right)
\left[\bigl(\partial_x w\bigr)^2 + \bigl(\partial_y w\bigr)^2 \right]  \partial_z w \right\} d x d y d z
\label{eqn:E3_sub_integral1}\\
 &= \left(\frac{1}{2}\, C_{33} + \frac{1}{6}\, C_{333} \right)
\int \frac{d^2 k}{(2\pi)^2} \frac{d^2 k'}{(2\pi)^2} \int\limits_{-\infty}^0 d z\,
\partial_z \tilde{w}(\mathbf{k}, z) \partial_z \tilde{w}(\mathbf{k}', z)
\partial_z \tilde{w}(\mathbf{k}'', z)
\label{eqn:E3_sub_integral2}\\
\mbox{} &-\left( \frac{1}{2}\, C_{13} + C_{44} + \frac{1}{2}\, C_{344}\right)
\int \frac{d^2 k}{(2\pi)^2} \frac{d^2 k'}{(2\pi)^2} \int\limits_{-\infty}^0 d z
\left(\mathbf{k} \cdot \mathbf{k}' \right)  
\tilde{w}(\mathbf{k}, z) \tilde{w}(\mathbf{k}', z) \partial_z \tilde{w}(\mathbf{k}'', z) \,,
\label{eqn:E3_sub_integral3}
\end{align}
where $\mathbf{k}'' = -\mathbf{k} - \mathbf{k}'$.
Substituting Eq.~\eqref{eqn:w_z} into Eqs.~\eqref{eqn:E3_sub_integral2},
\eqref{eqn:E3_sub_integral3}, and
symmetrizing with respect to $\mathbf{k}$, $\mathbf{k}'$, and $\mathbf{k}''$,
we obtain
\begin{equation}
\begin{aligned}
E^{(3)}_\mathrm{sub}
&= \left(\frac{1}{2}\, C_{33} + \frac{1}{6}\, C_{333} \right) \mu_2^2
\int \frac{d^2 k}{(2\pi)^2} \frac{d^2 k'}{(2\pi)^2}\,
\frac{k k' k''}{k + k' + k''}\,
\tilde{w}_0(\mathbf{k}) \tilde{w}_0(\mathbf{k}')
\tilde{w}_0(\mathbf{k}'')
\\
\mbox{} &+ \frac13 \left( \frac{1}{4}\, C_{13} + \frac{1}{2}\,C_{44} + \frac{1}{4}\, C_{344}\right)
\int \frac{d^2 k}{(2\pi)^2} \frac{d^2 k'}{(2\pi)^2}
\left(k^2 + {k'}^2 + {k''}^2
- 2\, \frac{k^3 + {k'}^3 + {k''}^3}{k + k' + k''} \right)
\tilde{w}_0(\mathbf{k}) \tilde{w}_0(\mathbf{k}') \tilde{w}_0(\mathbf{k}'') \,.
\label{eqn:F3}
\end{aligned}
\end{equation}
\end{widetext}
Setting $k = k' = k'' = Q$, we recover
the energy density postulated in the main text,
\begin{equation}
\mathcal{E}^{(3)}_\mathrm{sub} = \frac{1}{\Omega}\, E^{(3)}_\mathrm{sub}
= -\frac13 K_3 Q^2\, \left \langle w_0^3(x, y) \right\rangle ,
\label{eqn:F3_near_critical}
\end{equation}
with the coupling constant
\begin{equation}
\begin{split}
K_3  = &-\left(\frac{1}{2}\, C_{33} + \frac{1}{6}\, C_{333}\right)
\frac{C_{44}}{C_{33}}
\\
\mbox{} &- \left(\frac{1}{4}\, C_{13} + \frac{1}{2}\, C_{44} + \frac{1}{4}\, C_{344}\right)
\sim 20\,\mathrm{GPa}\,.
\end{split}
\label{eqn:K3}
\end{equation}
Here we have used theoretical estimates for graphite $C_{333} = -572\,\mathrm{GPa}$ and $C_{344} = -74.7\,\mathrm{GPa}$ ~\cite{Cousins2003}.
Note that the former is roughly consistent with the prediction $\frac12\, C_{33} + \frac16\, C_{333} = -\frac{17}{6}\, C_{33}$, i.e., $C_{333} = -20 C_{33} = -730\,\mathrm{GPa}$ of a simple model where graphite atoms interact
with atoms in adjacent layers via the Lennard-Jones potential $V(r) = {c_1}{r^{-12}} - {c_2}{r^{-6}}$.

The energy density correction $\mathcal{E}_{\mathrm{sub}}^{(3)}$
is easily computed for $n = 1, 2$, and $3$:
\begin{align}
\mathcal{E}^{(3)}_\mathrm{sub}(n = 1) &= \mathcal{E}^{(3)}_\mathrm{sub}(n = 2) = 0\,,
\\\
\mathcal{E}^{(3)}_\mathrm{sub}(n = 3) &= -\frac{1}{54} K_3 A_\mathrm{w}^3 Q^2 \cos \theta\,.
\label{eqn:E3_sub}
\end{align}
These results imply that the nonlinear correction favors the triangular pattern with the phase choice of $\theta = 0$, i.e., the pattern with the largest possible $\max w_0(x, y) = A_\mathrm{w}$ and $\min w_0(x, y) = -\frac13 A_\mathrm{w}$, over the stripe and the checkerboard patterns where $\max w_0 = -\min w_0 = A_\mathrm{w}$ and $E^{(3)}_\mathrm{sub}$ vanishes by symmetry.

\subsection{Film-substrate interaction}

The results of the previous sections enable us to formulate the
model for the film-substrate interaction.
Adding the linear and nonlinear elasticity contributions,
the interaction energy is
\begin{equation}
\begin{split}
E_{\mathrm{sub}} &= \frac12 \int \frac{d^2k}{(2\pi)^2} \tilde{\mathbf{u}}_0^\dagger(\mathbf{k}) \cdot \mathbf{B}_s(\mathbf{k}) \cdot \tilde{\mathbf{u}}_0(\mathbf{k})
\\[6pt]
\mbox{} &- \frac13\, K_3 Q^2 \int d x dy\, w_0^3(x, y) ,
\end{split}
\label{eqn:E_sub_full}
\end{equation}
Neglecting $u_0, v_0 \sim Q w_0^2 \ll w_0$, we get a simplified form~\cite{audolyboudaoud2008part1, Cai2011}
\begin{equation}
	E_{\text{sub}} =  \frac14 E^* \int \frac{d^2k}{(2\pi)^2}\,
    k\, |\tilde{w}(\mathbf{k})|^2 - \frac{1}{3} K_3 Q^2 \int d x d y\, w^3(x, y) \,,
\end{equation}
where $w$ refers to the surface deformation $(z = 0)$ from now on,
i.e., we drop the ``$0$'' subscript.
This is Eq.~(4) of the main text.

Taking the variational derivative of $E_{\text{sub}}$ with respect to $w$ gives the normal force exerted on the film by the substrate:
\begin{equation}
\begin{split}
	f_z(x, y) &= \frac12 E^* \int \frac{d^2k}{(2\pi)^2}\, k
    \tilde{w}(\mathbf{k})\, e^{i k_x x + i k_y y}
\\[6pt]
\mbox{} &- K_3 Q^2 w^2(x, y) .
\end{split}    
\end{equation}
Note that the first term is the standard Michell–Boussinesq response~\cite{Green1968} but with $E^*$ instead of $M$.

\begin{figure*}[t]
	\centering
	\includegraphics[width=0.8\textwidth]{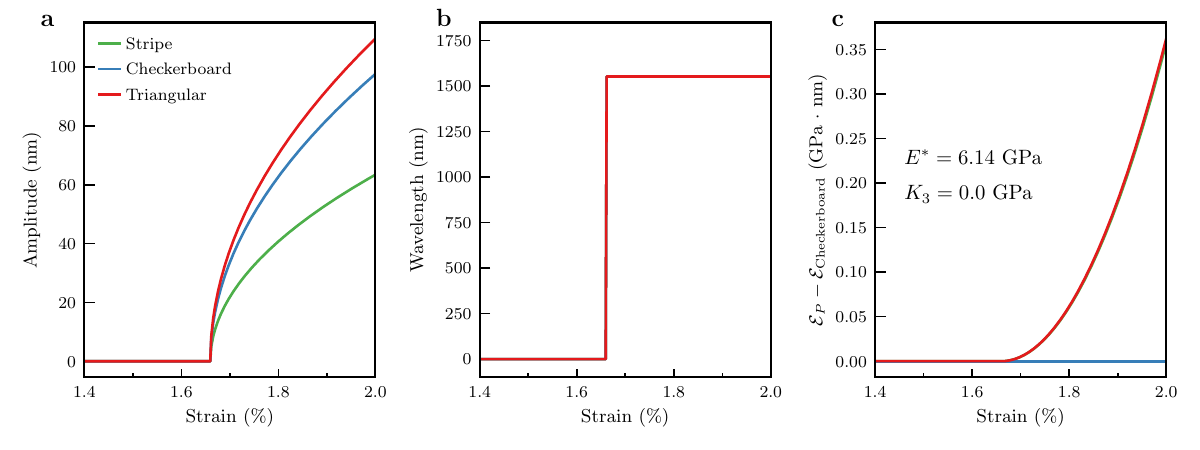}
	\caption{
		Solution of the FvK equations at $K_3 = 0$.
a. Wrinkle amplitude $A_\mathrm{w}$ as a function of external strain for
the stripe, checkerboard, and triangular patterns.
b. Wrinkle wavelength $\lambda_\mathrm{w}$ for each pattern.
c. Energy density of the stripe and triangular patterns relative to the checkerboard
pattern, which is the lowest energy state after the buckling.
Parameters: $E_\mathrm{f} = \SI{776}{GPa}$, $\nu = 0.208$, $E^* =\SI{6.14}{GPa}$, $h = \SI{70}{nm}$, $K_3 = 0$.
	}
\label{fig:FvK_pattern_comparison_1}
\end{figure*}

Since $f_z(x, y) = \sigma_{z z}\,|_{z = 0}$ plays a key role in the buckling phenomenon,
it is worthwhile to point out that Eq.~\eqref{eqn:sigma_xz_FvK}
of the FvK theory is approximate. A more accurate form of this equation,
which is
\begin{equation}
\sigma_{i z}(x, y, z) = \frac{h - z}{h}\, f_i(x, y)\,,
\qquad 0 < z < h\,,
\label{eqn:sigma_xz}
\end{equation}
includes the surface tractions in order to satisfy the correct boundary conditions at the top and bottom surfaces of the film, $z = 0$ and $z = h$.
The corresponding force balance equation is
\begin{equation}
	\partial_\beta \sigma_{\alpha\beta} = \frac{1}{h}\,
    f_\alpha(x, y)
\label{eqn:force_balance2}    
\end{equation}
instead of Eq.~\eqref{eqn:force_balance1}.
It implies that, strictly speaking, it is not possible to represent the in-plane strain
in the film in terms of an Airy function $\phi(x, y)$ in the manner of Eq.~\eqref{eqn:Airy_function}.
Instead, one should use
Eqs.~\eqref{eqn:E_FvK1}--\eqref{eqn:E_stretch1}
for the energy of the film and
\begin{equation}
\label{eqn:u_film}
u_{0,\alpha}(x, y) = u_{\alpha}\left(x,y\right) + \frac{h}{2}\, \partial_\alpha w\left(x,y\right)
\end{equation}
for the interface deformation $\mathbf{u}_0(x, y)$ in Eq.~\eqref{eqn:E_sub_full}.
This equation follows from the Kirchhoff approximation
\begin{align}
w(x, y, z) &=  w\left(x,y, \frac{h}{2}\right),
\label{eqn:w_film}\\
u_{\alpha}(x, y, z) &= u_{\alpha}\left(x,y, \frac{h}{2}\right)
    - \left(z - \frac{h}{2}\right) \partial_\alpha w\left(x,y, \frac{h}{2}\right)
\label{eqn:u_film}
\end{align}
that complies with Eq.~\eqref{eqn:e_xz_FvK}.
One can show that in this more refined approximation (which is similar to that
proposed in the Appendix of Ref.~\cite{Cai2011})
the critical strain is larger
than than $e_\mathrm{c}$ derived from the standard FvK theory
[Eq.~\eqref{eqn:e_c} in Sec.~\ref{sec:Trial}] by a small amount $\delta e_\mathrm{c} \sim e_\mathrm{c}^{3/2}$. 
However, this calculation is beyond the scope of the present work.

\section{Analysis of trial high-symmetry patterns}
\label{sec:Trial}

\begin{figure*}
	\centering
	\includegraphics[width=0.8\textwidth]{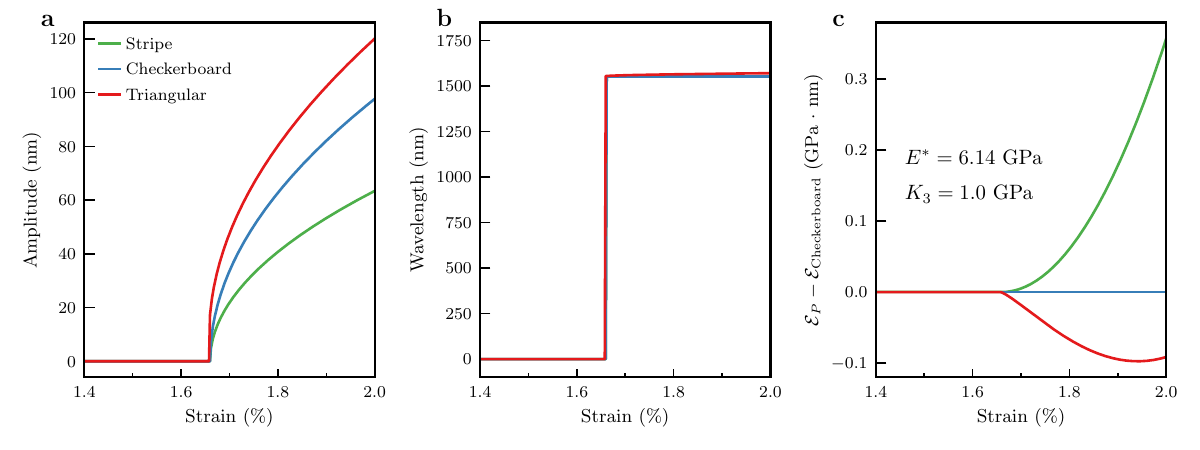}
	\caption{
Solution of the FvK equations at $K_3 \neq 0$.
a. Amplitude as a function of external strain for each pattern. 
b. Buckling wavelength for each pattern.
c. Energy density of each buckling pattern relative to the checkerboard pattern. After buckling the triangular pattern has the lowest energy.
Parameters are the same as in Fig.~\ref{fig:FvK_pattern_comparison_1} except $K_3 = \SI{1}{GPa}$.
}
\label{fig:FvK_pattern_comparison_2}
\end{figure*}

\begin{figure*}[!t]
\centering
\includegraphics[width=\textwidth]{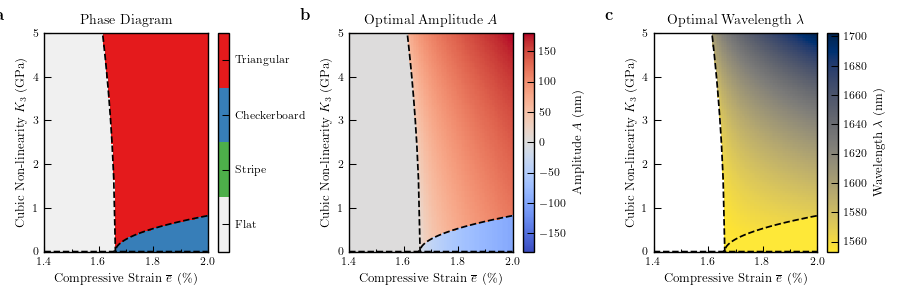}
\caption{
a. Buckling phase diagram as a function of the pre-strain $\bar{e}$ and the substrate nonlinearity $K_3$. Note that the stripe pattern is never favored.
The dashed lines are Eqs.~\eqref{eqn:K_3c1} and \eqref{eqn:K_3c2}.
b. Wrinkle amplitude of the lowest-energy pattern.
We chose $A < 0$ for the checkerboard pattern to show its phase boundary more clearly.
c. Wavelength of the optimal buckling pattern.
Parameters: $E_\mathrm{f} = \SI{776}{GPa}$, $\nu = 0.208$, $E^* =\SI{6.14}{GPa}$, $h = \SI{70}{nm}$,
same as in Figs.~\ref{fig:FvK_pattern_comparison_1} and \ref{fig:FvK_pattern_comparison_2}.
}
\label{fig:FvK_phase_diagram}
\end{figure*}

We evaluate the energy density $\mathcal{E} = \mathcal{E}_\mathrm{film} + \mathcal{E}_\mathrm{sub}$ of the three phases ---
the stripe phase ($C_{2z}$ symmetry), the checkerboard phase ($C_{4z}$ symmetry) and the triangular phase ($C_{6z}$ symmetry)
--- using the trial form of Eq.~\eqref{eqn:w0_triangular}
with $n = 1$, $2$, and $3$, respectively.
The calculation is simplified a bit by the fact that the mean strain tensor $\mathbf{E}$ [whose elements are given by Eq.~\eqref{eqn:average_strain}] is diagonal.
Solving for $\phi$ in Eq.~\eqref{eqn:Airy} and
substituting the result into Eq.~\eqref{eqn:E_film},
we obtain
\begin{align}
\mathcal{E}_\mathrm{film} &= \frac{1}{4 n}\, D_\mathrm{f} Q^4 A_\mathrm{w}^2
+\frac{1}{4 n} \,\frac{h E_\mathrm{f}}{1 - \nu} \left( -\bar{e} Q^2 A_\mathrm{w}^2
+ \frac{s_n}{64 n}\, Q^4 A_\mathrm{w}^4 \right),
\label{eqn:E_film_Audoly}\\
\mathcal{E}_\mathrm{sub} &= \frac{1}{8 n}\, E^* Q A_\mathrm{w}^2
+ \mathcal{E}_\mathrm{sub}^{(3)}\,.
\label{eqn:E_sub}
\end{align}
All the terms except for the nonlinear correction $\mathcal{E}_\mathrm{sub}^{(3)}$ [Eq.~\eqref{eqn:E3_sub}] have been derived previously~\cite{audolyboudaoud2008part1}.
Note however that our definition of $E^*$ differs from that in Ref.~\cite{audolyboudaoud2008part1}
by a factor of $4$. To lighten the notations we use $\nu$ in lieu of $\nu_\mathrm{H, f}$ again. 
The coefficient $s_n$ in Eq.~\eqref{eqn:E_film_Audoly} is $s_n = 2 / (1 + \nu)$ if $n = 1$ and
\begin{equation}
s_n = 1 + \left(\frac{3}{2} - \frac{2}{n}\right) (1 - \nu)
\label{eqn:s_n}
\end{equation}
otherwise. Equation~\eqref{eqn:s_n} seems to be valid not just for $n = 2$ and $3$ but for all $n > 1$ --- we checked it up to $n = 8$.

Introducing the rescaled wrinkle amplitude
\begin{equation}
a = A_\mathrm{w} / \sqrt{n}\,,
\end{equation}
the total energy density can be written as
\begin{equation}
\mathcal{E} = -\frac12\, \kappa_2 a^2 - \frac13 \kappa_3 a^3 + \frac14 \kappa_4 a^4 ,
\label{eqn:E_trial}
\end{equation}
with
\begin{align}
\kappa_2 &= -\frac{1}{2}\, \frac{h^3 E_\mathrm{f}}{12 (1 - \nu^2)} Q^4 + \frac12\, \frac{h E_\mathrm{f}}{1 - \nu}\, \bar{e} Q^2 - \frac{1}{4}\, E^* Q\,,
\label{eqn:kappa_2}\\
\kappa_3 &= \delta_{n, 3}\, \frac{\sqrt{3}}{6}\, K_3 Q^2\,,
\label{eqn:kappa_3}\\
\kappa_4 &= \frac{s_n}{64}\, \frac{h E_\mathrm{f}}{1 - \nu}\, Q^4\,.
\label{eqn:kappa4}
\end{align}
Note that $\kappa_2$ is the same for all the phases and that
$\kappa_3 \neq 0$ only for the triangular phase.
According to the variational principle, we can determine the upper bounds
on the critical strain $e_\mathrm{c}$ and the energy of the
post-buckled state at $\bar{e} > e_\mathrm{c}$ by minimizing the total energy of each phase with respect to $a$ and $Q$ at a given fixed $\bar{e}$.
For $n = 1$ and $2$, where the cubic term is absent,
the buckling occurs if $\kappa_2 > 0$. Its amplitude is
$a = \sqrt{\kappa_2 / \kappa_4}$ and its energy density is
\begin{equation}
\mathcal{E} = -\frac14\, \frac{\kappa_2^2}{\kappa_4}\,.
\label{eqn:E_cb}
\end{equation}

Since $\kappa_2$ is the same for all $n$, the post-buckled phase with the lowest energy is the one with the smallest $\kappa_4$, i.e., the smallest $s_n$.
It proves to be the checkerboard phase, $n = 2$~\cite{audolyboudaoud2008part1}.
The critical strain is found by minimizing the right-hand side of
Eq.~\eqref{eqn:E_cb} with respect to $Q$, which yields
\begin{equation}
Q = \frac{1}{h} \left(\frac{3 E^*}{\overline{E}_\mathrm{f}} \right)^{1/3}
\label{eqn:Q}
\end{equation}
independent of $\bar{e}$.
The corresponding wrinkle period is $\lambda_\mathrm{w} = 2\pi / Q$.
The critical strain is obtained by solving $\kappa_2 = 0$
for $\bar{e}$, which gives
\begin{equation}
e_\mathrm{c} = \frac14\, \frac{1}{1 + \nu}
\left( \frac{3 E^*}{\overline{E}_\mathrm{f}} \right)^{2/3} .
\label{eqn:e_c}
\end{equation}
Above the critical point, the wrinkle amplitude in the
checkerboard phase grows as
\begin{equation}
\begin{split}
A_\mathrm{w}\left(n = 2\right) &= \sqrt{2}\, a
= c_0 h
\left(\frac{\bar{e}}{e_\mathrm{c}} - 1\right)^{1/2} ,
\\
c_0 &= \sqrt{\frac{32}{(1 + \nu)(3 - \nu)}}\,.
\end{split}
\label{eqn:A_cb}
\end{equation}
\smallskip

Such a behavior is typical for a second-order phase transition
in the Landau theory.

For the triangular phase, the transition becomes
of the first order because of the cubic term in the
energy. This transition occurs when
\begin{equation}
\kappa_2 = -\frac{2}{9}\, \frac{\kappa_3^2}{\kappa_4} < 0
\qquad (n = 3)\,.
\label{eqn:kappa_3c1}
\end{equation}
The corresponding $K_3$  as a function of $\bar{e}$ is
\begin{equation}
K_3 = c_1 E^* \frac{\sqrt{e_\mathrm{c} - \bar{e}}}{e_\mathrm{c}}\,,
\quad c_1 = \frac{9}{32} \left(\frac{11 - 5 \nu}{2}\right)^{1/2} .
\label{eqn:K_3c1}
\end{equation}
Inverting this, we get the threshold strain as a function of $K_3$:
\begin{equation}
\label{eqn:critical_buckling_K3}
    e_c^{\,\ast}(K_3) = e_c - \left(\frac{e_c}{c_1}\frac{K_3}{E^*}\right)^{2} .
\end{equation}
\begin{widetext}
As $\bar{e}$ increases, the triangular phase eventually looses the
competition to the checkerboard phase at
\begin{equation}
\kappa_2 = \frac{9 \rho - \rho^2 + (3 + \rho)^{3/2} \rho^{1/2}}{9 (1 - \rho)^2}\,
\frac{\kappa_3^2}{\kappa_4} > 0\,,
\quad
\rho = \frac{s_2}{s_3} \qquad(n = 3)
\,.
\label{eqn:kappa_3c2}
\end{equation}
The corresponding $K_3$ is
\begin{equation}
\begin{split}
K_3 = c_2 E^* \frac{\sqrt{\bar{e} - e_\mathrm{c}}}{e_\mathrm{c}}\,,
\qquad
c_2 = \frac{3}{32}
\left[\frac{2 (11 - 5 \nu) (1 - \nu)^2}{(15 - 7 \nu) (3 - \nu) + \sqrt{6 - 2\nu}\, (7 - 3\nu)^{3/2}}\right]^{1/2} .
\label{eqn:K_3c2}
\end{split}
\end{equation}
To verify these analytical results we carried out numerical minimization of the energy given by Eq.~\eqref{eqn:E_trial}
with respect to $a$ and $Q$.
Figure~\ref{fig:FvK_pattern_comparison_1} compares the
optimized energies of the three phases as a function of $\bar{e}$ at $K_3 = 0$. Each buckling pattern has the same threshold $e_\mathrm{c}$ and the checkerboard pattern has the lowest energy after buckling,
as expected.
Figure~\ref{fig:FvK_pattern_comparison_2} shows the energies of these phases at a nonzero $K_3$. In this case the buckling into the triangular phase with a finite amplitude $A_\mathrm{w}$
occurs abruptly at a smaller $\bar{e}$ than for the other two phases,
in agreement with Eq.~\eqref{eqn:critical_buckling_K3}.
As the strain increases,
the period of the triangular phase grows slightly with $\bar{e}$.
This phase looses
the energy competition to the checkerboard one when $\bar{e}$ exceeds
another threshold. 
In Fig.~\ref{fig:FvK_phase_diagram}a (same as Fig.~1a of the main text) we show the phase diagram obtained from these calculations and in Figs.~\ref{fig:FvK_phase_diagram}b, \ref{fig:FvK_phase_diagram}c we present our results for the optimal amplitude $A_\mathrm{w} = \max |w(x, y)|$ and the wavelength
$\lambda_\mathrm{w}$, respectively.

\end{widetext}

\clearpage 